\documentclass[%
 reprint,
superscriptaddress,
 amsmath,amssymb,
 aps,
]{revtex4-2}

\usepackage{graphicx}% Include figure files
\usepackage{dcolumn}% Align table columns on decimal point
\usepackage{bm}% bold math
\usepackage{color}
\usepackage{qcircuit}
\usepackage{booktabs,eqparbox}

\usepackage[utf8]{inputenc}
\usepackage{pgfplots}
\usepackage{hyperref}

\DeclareUnicodeCharacter{2212}{−}
\DeclareMathOperator{\diag}{diag}
\usepgfplotslibrary{groupplots,dateplot}
\usetikzlibrary{patterns,shapes.arrows}
\pgfplotsset{compat=newest}

\begin{document}

\title{Quantum gates with parametrically driven multi-qubit couplers}

\author{Verena Feulner}
\email{verena.vf.feulner@fau.de}
\affiliation{Physics Department, Friedrich-Alexander-Universität Erlangen Nürnberg, Germany}

\author{Marjan Fani}
\affiliation{Physics Department, Friedrich-Alexander-Universität Erlangen Nürnberg, Germany}

\author{Lukas Heunisch}
\affiliation{Physics Department, Friedrich-Alexander-Universität Erlangen Nürnberg, Germany}

\author{Stephan Tasler}
\affiliation{Physics Department, Friedrich-Alexander-Universität Erlangen Nürnberg, Germany}

\author{Michael J. Hartmann}
 \email{michael.j.hartmann@fau.de}
\affiliation{Physics Department, Friedrich-Alexander-Universität Erlangen Nürnberg, Germany}
\affiliation{Quint Computing GmbH, Erwin-Rommel-Str. 1, 91058 Erlangen, Germany}

\date{\today}

\begin{abstract}
Superconducting quantum processors could significantly profit from enhanced connectivity together with precise control of interactions and gates between qubits. Here we investigate plaquettes of four qubits that are coupled via a central tunable coupling circuit, so that not only gates between qubits connected by an edge of the plaquette can be executed but also between qubits across the diagonal. By numerically and analytically analyzing parametrically driven processes, we explore $\sqrt{\text{iSWAP}}$-gates between any pair of qubits, also across the diagonal, as well as three-qubit interactions and gates. For experimentally available circuit parameters, we for example find $\sqrt{\text{iSWAP}}$-gates with a gate time of 50 ns and 99.9\% fidelity, which is decreased to 99.4\% if two such gates are executed in parallel on disjoint qubit pairs in the plaquette. For three-qubit gates we find fidelities of 95\% fidelity at a gate time of 200 ns.

\end{abstract}

\maketitle

\section{\label{sec:intro} Introduction}
Superconducting qubits have emerged as one of the leading platforms for quantum computing, because of their compatibility with existing microwave electronics, scalability and fast gate times. Among various implementations for qubits, the transmon qubit \cite{PhysRevA.76.042319} is the most widely used due to its reduced sensitivity to charge noise and ease of fabrication.
 
In the superconducting architectures used today, qubit-qubit interactions are typically mediated by tunable couplers \cite{Campbell_2023, Sameti_2019}, which offer the ability to suppress residual interactions and dynamically configure the connectivity of the quantum chip.
A prominent technique to control such couplings, is by parametric modulation of the coupler \cite{PhysRevApplied.19.044003, PhysRevX.11.021058, PhysRevApplied.6.064007, PhysRevLett.113.220502, articleNakamuratunable,Campbell_2023, Sameti_2019}. For example, applying a microwave-drive at a frequency that matches the detuning between the connected qubits, can selectively turn interactions on and off. This enables the realization of fast, high-fidelity gates, which are essential for the execution of quantum algorithms.

Despite the success of current two-qubit coupler schemes, the limited connectivity of superconducting quantum processors remains a critical challenge \cite{PhysRevApplied.16.024018}. This often means that deeper circuits with more gates are needed than in high connectivity platforms, which in turn limits the applications that are feasible. In the context of quantum error correction, low connectivity restricts the set of error correcting codes that could be implemented in a device or means that stabilizer elements need to be read out via longer sequences of two-qubit gates that map correlations of data qubits onto a measurement qubit \cite{Google_scaling_log_qubit}.  For instance, many families of quantum low-density parity-check (qLDPC) codes \cite{PRXQuantum.2.040101}, involve stabilizers on qubits that are not nearest-neighbor on a 2D-qubit grid, making them difficult to realize on planar superconducting qubit chips \cite{PRXQuantum.4.020321}. These codes offer higher encoding rates and a possibly lower overhead. 

\begin{figure}[h!]
    \centering
    \includegraphics[width = 0.48\textwidth]{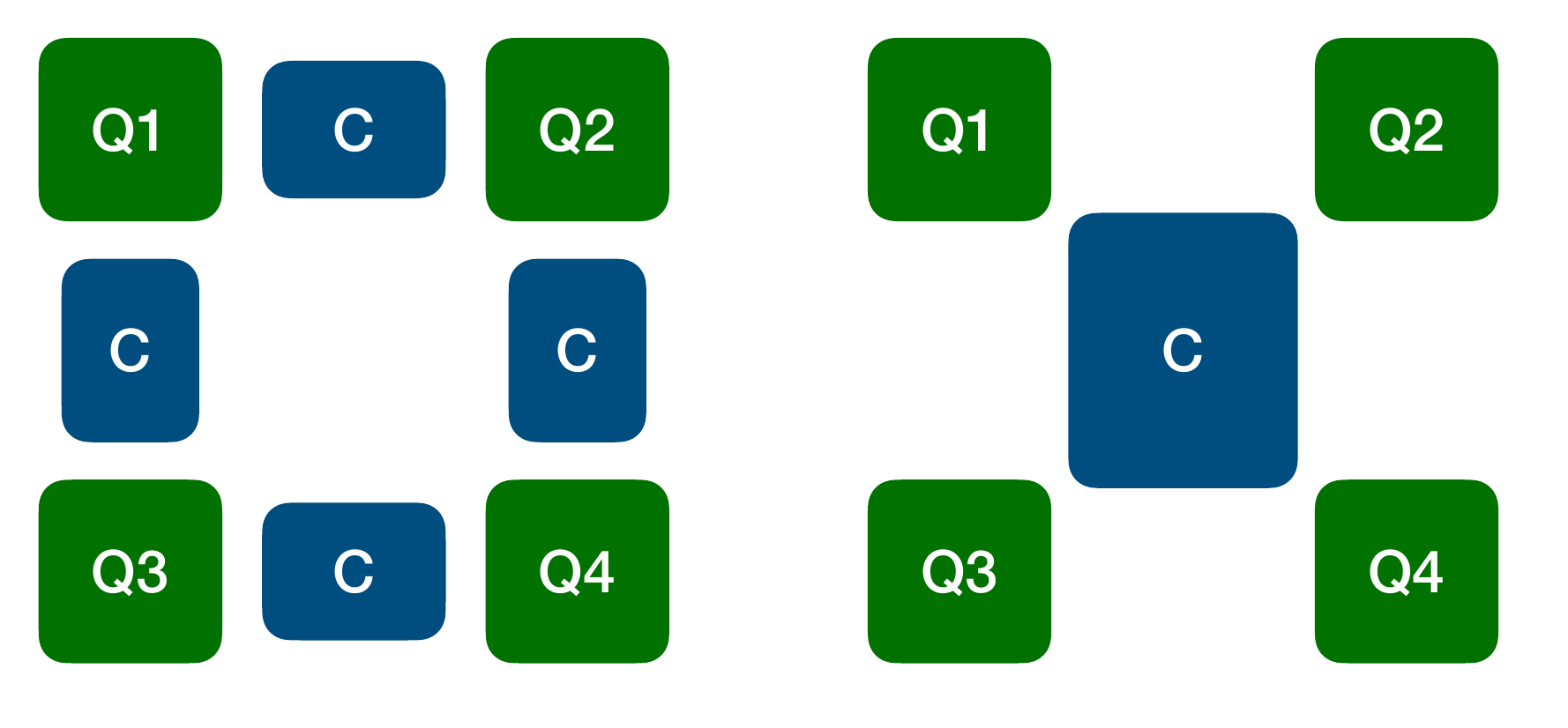}
    \includegraphics[width = 0.48\textwidth]{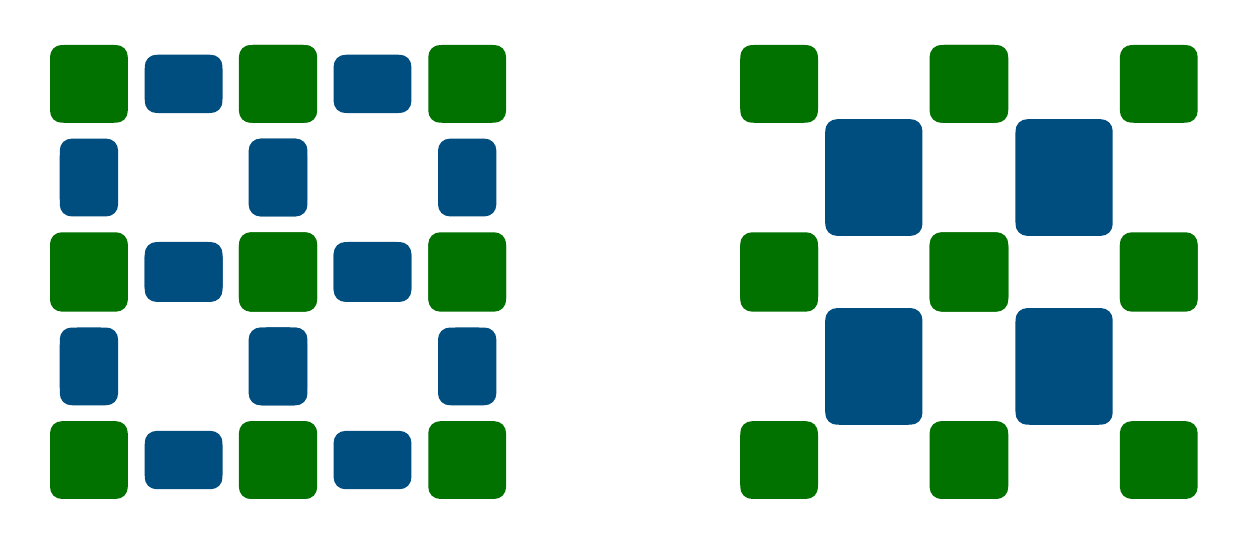}
		    \caption{Comparison of rectangular 4-qubit modules. Top: On the left the qubits are connected by four two-qubit couplers and one multi-qubit coupler positioned in the middle of these qubits (right). Bottom: Arrangements with 2-qubit couplers or 4-qubit couplers in larger grids.}
            \label{fig:chip design}
\end{figure}

To address these challenges, multi-qubit couplers are being investigated as a promising solution \cite{Sameti_2017,Miyazaki_2025, zhou_realizing_2023,kawakami2025fourbodyinteractionskerrparametric} since
they provide higher effective connectivity within the same physical layout. In standard two-qubit coupler architectures, interactions are restricted to directly adjacent qubits. Operations between non-adjacent qubits, such as pairs across the diagonal of a square unit cell, require state-routing via SWAP networks. For example, implementing an iSWAP gate between two diagonally separated qubits in a rectangular grid typically requires three two-qubit executions (two SWAPs plus the target iSWAP gate), thus introducing further errors with every gate, see Fig. \ref{fig:chip design}. 

Parametric driving techniques in combination with a multi-qubit coupler overcome this limitation by enabling a single coupler module to mediate tunable interactions among multiple, mutually detuned qubits. By selectively activating frequency-matched interactions, the system can realize directly the targeted interactions between non-adjacent qubits without the need for SWAP-based routing. 

These devices can also reduce the total number of circuit components by enabling couplings between more than two qubits with a single coupler. For example, in a typical 2D rectangular architecture with $m$ rows and $n$ columns of qubits, one requires $m(n-1)+n(m-1)$ standard two-qubit couplers. In contrast, introducing four-qubit couplers which are placed at the center of each 2x2 qubit cell, requires only $(m-1)(n-1)$ couplers, so a factor 2 less couplers for $n, m \gg 1$ see Fig. \ref{fig:chip design}.

Importantly, multi-qubit couplers can mediate, not only pairwise, but higher-order interactions involving three or four qubits simultaneously. Such interactions are of particular interest for quantum simulation, for instance for simulating models with intrinsic multi-body terms like the Kitaev model or lattice gauge theories \cite{busnaina2025nativethreebodyinteractionssuperconducting, Banuls_2020, PhysRevResearch.5.023077}. They can also be helpful for implementing quantum error-correcting codes in a more compact way by running multi-body parity checks in a single step \cite{tasler2025optimizingsuperconductingthreequbitgates, christensen2023schemeparitycontrolledmultiqubitgates}.

However, the introduction of multi-qubit couplers also brings new challenges. The shared coupling element increases the hybridization and crosstalk between the modes, as multiple qubits interact through a common element. This can complicate the isolation and control of individual gates and may degrade the overall fidelity of the processor.

In this work, we investigate two candidate implementations for tunable multi-qubit couplers: a modified Superconducting Nonlinear Asymmetric Inductive eLement (SNAIL) and a dc SQUID-based design. We simulate parametrically driven gates between pairs of qubits and evaluate their performance in terms of fidelity. Furthermore, we demonstrate the possibility of simultaneously activating multiple interactions. Enabling, for example, two 2-qubit gates in parallel. We also explore the potential for higher-order gate operations.
Finally, we assess the scalability of these coupler designs and discuss technical challenges in improving the performance and in the integration into large-scale quantum processors.

\section{\label{sec:theory} Circuit Architectures}

We start by describing the two circuit architectures that we consider for multi-qubit coupler circuits. These are a Superconducting Quantum Interference Device (dc-SQUID) and a Superconducting Nonlinear Asymmetric Inductive eLement (SNAIL). We start with the dc-SQUID.

\subsection{dc-SQUID coupler}
\begin{figure}[h]
		\begin{center}
			\includegraphics[width = 0.25\textwidth]{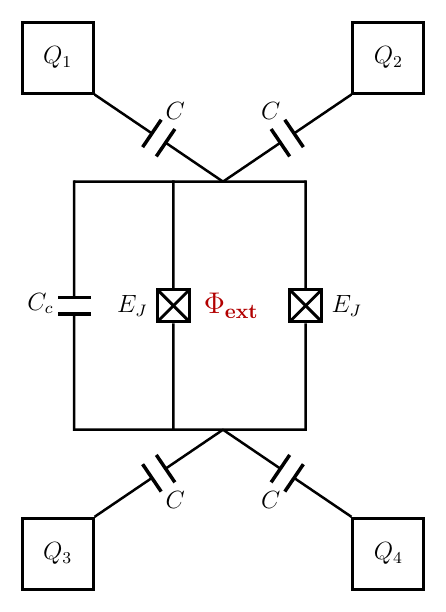}
		    \caption{Layout of a dc-SQUID as a multi-qubit coupler capacitively coupled to four transmon qubits. $C$ are the coupling capacitances between the coupler and the qubits (Q1-Q4). The SQUID loop consists two junctions in parallel with Josephson energies $E_J$ each. The SQUID is shunted by a capacitance $C_c$. The flux-loop of the SQUID can be tuned by an external flux $\Phi_{\text{ext}}$.}
            \label{fig:design dc-SQUID}
		\end{center}
		
	\end{figure}

The circuit  of a dc-SQUID (Superconducting Quantum Interference Device), acting as a coupler that connects four qubits, is shown in Fig. \ref{fig:design dc-SQUID}. The four transmon qubits are capacitively coupled to a dc-SQUID, which is modulated by an external time-dependent flux. A SQUID is a superconducting loop with two Josephson junctions. Its effective Josephson energy $E_J(\phi_{\text{ext}})$ depends periodically on the external magnetic flux threaded through the loop. Therefore, the frequency becomes tunable and by placing it between qubits the interactions can be controlled by the external flux \cite{fay2010quantumdynamicsdcsquidcoupled,PhysRevA.96.062323}.
 The parameters that we assume for the circuit are given in table \ref{table:table_squid}. 

\begin{table}[h]
\centering
\begin{tabular}{ |c|c|c| } 
 \hline
  & Frequency [GHz] & Anharmonicity [GHz] \\ 
  \hline
Qubit 1 & 7.12 & -0.191 \\ 
Qubit 2 & 6.578 & -0.191 \\ 
Qubit 3 & 6.171 & -0.191 \\ 
Qubit 4 & 5.539 & -0.191 \\ 
SQUID   & 13.207 & -0.168 \\ 
 \hline
\end{tabular}

\caption{Considered transition frequencies and anharmonicities for the circuit with the dc-SQUID-coupler. These parameter values are a consequences of the chosen capacitances and Josephson energies, see table \ref{table:table_squid_capE}, where capacitance values for the transmons are taken from \cite{RevModPhys.93.025005} as an example. }
\label{table:table_squid}
\end{table}

The circuit in Fig. \ref{fig:design dc-SQUID} can be quantized using the standard approach in cQED leading to a system Hamiltonian,
\begin{align} \label{eq:ham}
    H(t) = \sum_{j=1}^4 H_j +H_c(t) +H_g,
\end{align}
where the four transmon qubits are described by Hamiltonians
\begin{align}
    H_j = 4E_{Cj}n_j^2 -E_{Jj}\cos(\varphi_j),
    \label{eq:transmonsHam}
\end{align}
and the dc-SQUID by
\begin{align}
\begin{aligned}
H_c(t) = 4E_{Cc}n_c^2 &
- E_{Jl}\cos\left(\varphi_c+m_l\varphi_{\text{ext}}(t)\right) 
\\&-E_{Jr}\cos\left(\varphi_c+m_r\varphi_{\text{ext}}(t)\right) .
\label{eq:SQUID_Ham_full}
\end{aligned}
\end{align}
Here, the $\varphi_i$ and $\varphi_c$ are the superconducting phase degrees of freedom with their conjugate Cooper-pair numbers $n_i$ and $n_c$, where we take $\hbar=1$. The Josephson energies are denoted by $E_{Ji}$ for the transmons and by $E_{Jr}$ ($E_{Jl}$) for the right (left) junction of the coupler-loop.
The numbers $m_r$ and $m_l$ fix the gauge of the description and need to be chosen carefully due to the time-dependence of the drive \cite{irrot_gauge}, as will be discussed in more detail in section \ref{sec:tdep}.
In this paper we consider a symmetric SQUID $(E_{Jl}=E_{Jr}=E_J)$, where the Hamiltonian \ref{eq:SQUID_Ham_full} can be written as,
\begin{align}
\begin{aligned}
    H_c(t) = 4E_{Cc}n_c^2  &- E_{J}\cos(\varphi_c+\frac{1}{2}\varphi_{\text{ext}}(t)) \\&- E_{J}\cos(\varphi_c-\frac{1}{2}\varphi_{\text{ext}}(t)),\end{aligned}
\end{align}
since, due to the symmetry of the junctions, $m_l = 1/2$ and $m_r = -1/2$.
The system is driven by threading the coupler loop by an external flux $\varphi_{\text{ext}}$, that consists of a constant part $\varphi_{\text{dc}}$ and a time-dependent part $\varphi_{\text{ac}}$,
\begin{align}
    \varphi_{\text{ext}}(t) = \varphi_{\text{dc}}+\varphi_{\text{ac}}(t).
\end{align}
The time-dependent part is modulated by a cosine pulse with a modulation amplitude $\delta$,
\begin{align}
    \varphi_{\text{ac}}(t) = \delta\cos(\omega_D t),
    \label{eq:phiAc}
\end{align}
where $\omega_D$ is the drive frequency of the parametric drive.

The coupling Hamiltonian arises due to the capacitive coupling of the qubits to the SQUID and the mutual capacitive coupling between the qubits,
\begin{align}
    H_g =\sum_{j=1}^4 4E_{Cjc}n_jn_c +\sum_{i,j=1, i\neq j}^4 4E_{Cij}n_in_j
    \label{eq:couplingHam}
\end{align}
 $E_{Cjc}$ and $E_{Cij}$ are the entries of the charging energy matrix of the coupler and the qubits. 

 In terms of bosonic creation and annihilation operators $a_i^\dagger$ and $a_i$ with $[a_i,a_i^\dagger ]=1$($i = 1,2,3,4, c$), the charge and flux operators read,
\begin{align*}
    n_i & = \frac{i}{\sqrt{2}}\left(\frac{E_{J_i}}{8E_{C_i}}\right)^{\frac{1}{4}}(a_i^\dagger-a_i),\\
    \varphi_i & = \frac{1}{\sqrt{2}}\left(\frac{8E_{C_i}}{E_{J_i}}\right)^{\frac{1}{4}}(a_i^\dagger+a_i).
\end{align*}
and the dc-SQUID system can be described as
\begin{align}
\begin{aligned}
    &H_t= \sum_{i=1}^4 \omega_i a_i^\dagger a_i + \frac{\alpha_i}{2} a_i^\dagger a_i^\dagger a_i a_i, \\
  &H_c(t) =\omega_c(t) a_c^\dagger a_c + \frac{\alpha_c(t)}{2} a_c^\dagger a_c^\dagger a_c a_c ,\\
  & H_g = \sum_{i=1}^4 g_{ic} \left(a_i^\dagger a_c + a_c^\dagger a_i\right) +\sum_{i,j=1, i\neq j}^4 g_{ij}\left(a_i^\dagger a_j + a_j^\dagger a_i\right) .
\end{aligned} 
\end{align}
where we have expanded the potentials of the qubits and coupler to 4th order in the respective phase variables. $\omega_i=\sqrt{8E_{Ci}E_{Ji}}-E_{Ci}$ and $\alpha_i = -E_{Ci}$. $g_{ic}$ is the qubit-coupler coupling, $g_{ij}$ are the qubit-qubit couplings 

$\omega_c=\sqrt{8E_{Cc}E_{J}(t)} -\frac{1}{2}\varphi_{\text{zpf}}^4E_{J}(t)$ and $\alpha_c(t) = -\frac{1}{2}\varphi_{\text{zpf}}^4E_{J}(t)$. 

\subsection{SNAIL-coupler}

\begin{figure}[h]
		\begin{center}
			\includegraphics[width = 0.25\textwidth]{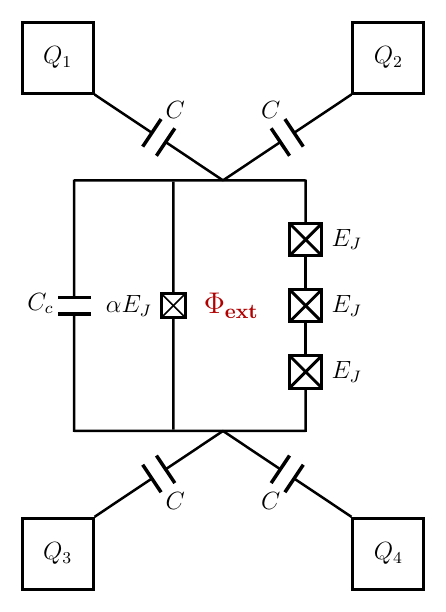}
		    \caption{Layout of a SNAIL as a multi-qubit coupler capacitively coupled to four transmon qubits. $C$ are the coupling capacitances between the coupler and the qubits (Q1-Q4). The SNAIL loop consists of three big junctions with a Josephson energies $E_J$ each  and a smaller Josephson junction with energy $\alpha E_J$. The SNAIL is shunted by a capacitance $C_c$. The flux-loop of the SNAIL can be tuned by an external flux $\Phi_{\text{ext}}$.}
            \label{fig:design SNAIL} 
		\end{center}
	\end{figure}

The other option we consider to couple four qubits, is the circuit shown in Fig. \ref{fig:design SNAIL}. It consists of the four considered transmon qubits that are here capacitively coupled to a SNAIL circuit, which is modulated by an external time-dependent flux. The SNAIL consists of a superconducting loop that is interrupted by an asymmetric array of Josephson junctions. Typically several larger junctions with Josephson energy $E_J$ (often three) and one smaller junction with Josephson energy $\alpha E_J$, whose strength is determined by the parameter $\alpha < 1$ are used. The asymmetry of these junctions gives rise to nonlinearities of odd order, in particular a third-order nonlinearity -- $\varphi^3$ \cite{Frattini_2017, Frattini_2018,Sivak_2019}. The effective potential of the SNAIL depends on the external magnetic flux threading the loop, which allows for a flux-tunable frequency and nonlinearity.
The parameters that we choose for our SNAIL circuit are given in table \ref{table:table_snail}. 
\begin{table}[h]
\centering
\begin{tabular}{ |c|c|c| } 
 \hline
  & Frequency [GHz] & Anharmonicity [GHz] \\ 
  \hline
Qubit 1 & 6.940 & -0.208 \\ 
Qubit 2 & 6.697 & -0.208 \\ 
Qubit 3 & 6.314 & -0.208 \\ 
Qubit 4 & 5.764 & -0.208 \\ 
SNAIL   & 12.927 & -0.261 \\ 
 \hline
\end{tabular}
\caption{Considered transition frequencies and anharmonicities for the circuit with the SNAIL-coupler. These parameter values are a consequences of the chosen capacitances and Josephson energies, see table \ref{table:table_snail_capE}, where approximate capacitance values for the transmons are taken from \cite{RevModPhys.93.025005} as an example. }
\label{table:table_snail}
\end{table}

The circuit shown in Fig. \ref{fig:design SNAIL} can be described by a Hamiltonian that is of the same form as in Eq. (\ref{eq:ham}), where the four transmon qubits are again described by Hamiltonians as in Eq. (\ref{eq:transmonsHam}), but for which the coupler is now described by a SNAIL Hamiltonian,
\begin{align}
\begin{aligned}
H_c(t) = 4E_{Cc}n_c^2 &
- \alpha E_J\cos\left(\varphi_c+m_l\varphi_{\text{ext}}(t)\right) 
\\&-3 E_J\cos\left(\varphi_c/3+(m_r/3) \varphi_{\text{ext}}(t)\right),
\label{eq:SNAIL_Ham_full}
\end{aligned}
\end{align}
with Josephson energies $E_J$ for three big junctions, $\alpha E_J$ for the small junction and charging energy $E_{Cc}$.
The coupling Hamiltonian $H_g$ that arises due to the capacitive coupling of the qubits to the SNAIL is the same as for the SQUID , see Eq. (\ref{eq:couplingHam}).

\begin{figure}[h]
    \centering
    \begin{minipage}{0.5\textwidth}
        \centering
        \includegraphics[width = 0.5\textwidth]{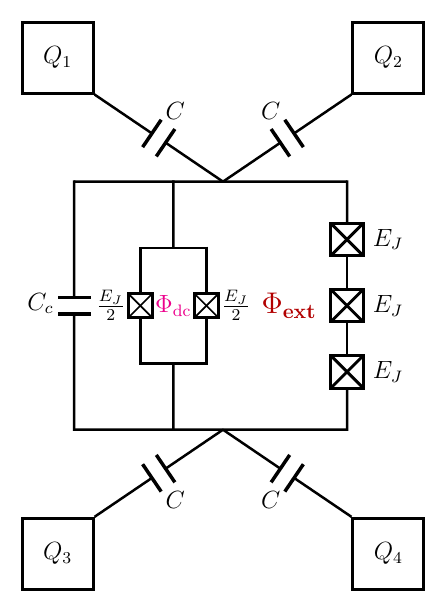}
		    \caption{To access the different regimes of the SNAIL, depending on the parameter ratio $\alpha$ of the smaller junction, we make the $\alpha$ tunable by using a SQUID-loop, that can be tuned by a time-independent pulse to tune the $\alpha$ from 0 to 1.}
            \label{fig:tunable SNAIL circuit}
    \end{minipage}
    \begin{minipage}{0.5\textwidth}
        \centering
        \includegraphics[width = 0.9\textwidth]{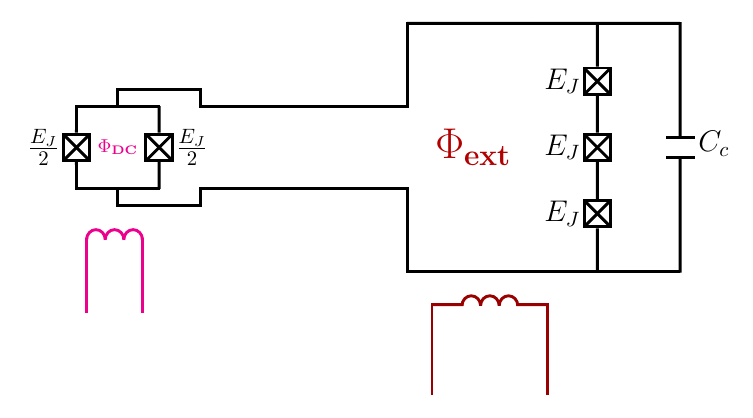}
		    \caption{More realistic layout of the tunable SNAIL. Both flux loops should have a spacial separation to avoid flux-crosstalk.}
            	\label{fig:tunable SNAIL design}
    \end{minipage}

	\end{figure}

To make our approach fully versatile, it is interesting to design it such that one can switch between different interesting regimes in one device on the fly. To be able to do so, we consider a SNAIL with a tunable $\alpha$-junction and 3 big Josephson junctions. This is useful for switching between regimes, where a gate for two qubits can be done fast with a large $\alpha$-value, and a regime with a small $\alpha$ and therefore a high nonlinearity, for three- and four-qubit gates. 
The circuit for a SNAIL with tunable $\alpha$-value is shown in Fig. \ref{fig:tunable SNAIL circuit}. Here, the SNAIL can be driven by an external flux through the big loop to drive gates, while the value of $\alpha$ can be tuned by applying a dc-pulse to the small SQUID-loop. 

Flux crosstalk can be prevented in this case by fabricating the both loops of the coupler spatially separated  as shown in Fig. \ref{fig:tunable SNAIL design}.

\subsection{Analysis of the time-dependent external drives\label{sec:tdep}}

The couplers are driven with a time-varying pulse, which requires a careful choice of the gauge in terms of the weighting factors $m_l$ and $m_r$ in Eq. \ref{eq:SQUID_Ham_full} and \ref{eq:SNAIL_Ham_full}. For transparency and ease of interpretation, we choose the so-called irrotational gauge \cite{irrot_gauge}. This choice fixes the gauge via the irrotational constraint $C_r m_l +C_l m_r = 0$, resulting in \footnote{As in \cite{irrot_gauge}, this analysis is already done at the level of the Lagrangian formalism, where e.g. for a SNAIL, $\mathcal{L}_{\text{SNAIL}} = \frac{C_\Sigma}{2}\left(\frac{\Phi_0}{2\pi}\right)^2\dot{{\varphi}}^2 + 3 E_J\cos\left(\frac{\varphi-m_r\varphi_e}{3}\right) +\alpha E_J \cos\left(\varphi +m_l\varphi_e\right)$.},
\begin{align*}
    m_r = -\frac{C_l}{C_r+C_l}, \hspace{0.3cm} m_l=\frac{C_r}{C_r+C_l} \, . \hspace{0.3cm} 
\end{align*}
Where the $C_l$ and $C_r$ are proportional to the factor $\alpha$ and 3 of the SNAIL junctions due to the proportionality of both $E_J$ and $C$ to the junction area.

\subsubsection{Analysis of the SNAIL-drive via the Jacobi-Anger expansion \label{sec:Jacobi-Anger}}
To further analyze the effects generated by the ac-drive, we make use of the Jacobi-Anger expansion \cite{abramowitz+stegun}, which is a generalized Fourier series of the external drive terms. This expansion is useful for investigating the harmonic content of periodic phenomena \cite{DATTOLI1996183}. 

To describe the periodic cosine and sine terms of the drive, we thus use the real-valued variations of the Jacobi-Anger expansion,
\begin{align}
\cos(z\cos(\theta)) &= J_0(z) +2\sum_{n=1}^\infty (-1)^n J_{2n}(z)\cos(2n\theta), 
\label{eq:JA_cos} \\
     \sin(z\cos(\theta)) &= -2\sum_{n=1}^\infty (-1)^n J_{2n-1}(z)\cos((2n-1)\theta),
     \label{eq:JA_sin}
\end{align}
where the $J_n(z)$ are the nth Bessel functions of the first kind.

Whereas the external flux $\varphi_{\text{ext}}=\varphi_{\text{dc}}+\varphi_{\text{ac}}$, generally contains a time-independent part $\varphi_{\text{dc}}$ and a time-dependent part  $\varphi_{\text{ac}}(t) = \delta\cos(\omega_D t)$, see Eq. \ref{eq:phiAc},  we here find that the interactions we are interested in can be generated with $\varphi_{\text{dc}}=0$ and make this choice for simplicity. Making 
use of trigonometric identities and the expansions in Eqs. (\ref{eq:JA_cos}) and (\ref{eq:JA_sin}), we find for the two contributions to the potential in Eq. (\ref{eq:SNAIL_Ham_full}),

\begin{widetext}
\begin{align}
\begin{aligned}
 \cos\left(\frac{\varphi_c}{3}+\frac{m_r}{3}\delta\cos(\omega_D t)\right)&=\cos\left(\frac{\varphi_c}{3}\right)\left(J_0\left(\frac{m_r\delta}{3}\right) +2\sum_{n=1}^\infty (-1)^n J_{2n}\left(\frac{m_r\delta}{3}\right)\cos(2n\omega_Dt)\right) \\
 &+\sin\left(\frac{\varphi_c}{3}\right)\left(-2\sum_{n=1}^\infty (-1)^n J_{2n-1}\left(\frac{m_r\delta}{3}\right)\cos((2n-1)\omega_Dt)\right),
\end{aligned}
\label{SNAIL_1st_term_JA}
\end{align}
and 
\begin{align}
    \begin{aligned}
       \cos\left(\varphi_c+m_l\delta\cos(\omega_D t)\right)&=\cos(\varphi_c)\left(J_0(m_l\delta) +2\sum_{n=1}^\infty (-1)^n J_{2n}(m_l\delta)\cos(2n\omega_Dt)\right) \\
 &+\sin(\varphi_c)\left(-2\sum_{n=1}^\infty (-1)^n J_{2n-1}(m_l \delta)\cos((2n-1)\omega_Dt)\right).
    \end{aligned}
\label{SNAIL_2nd_term_JA}    
\end{align}
\end{widetext}

\begin{figure}
		\begin{center}
			\includegraphics[width = 0.45\textwidth]{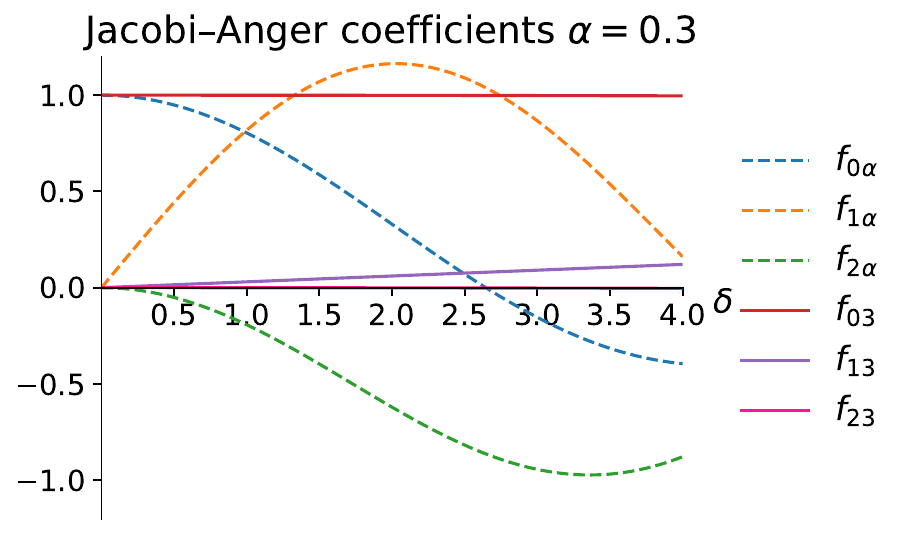}
		    \caption{Magnitude of the different Bessel-function components (without the $\cos$-time-dependence) for the SNAIL depending on the drive amplitude $\delta$ for $\alpha=0.3$. }
            \label{fig:bessel1}
		\end{center}
\end{figure}

\begin{figure}
		\begin{center}
			\includegraphics[width = 0.45\textwidth]{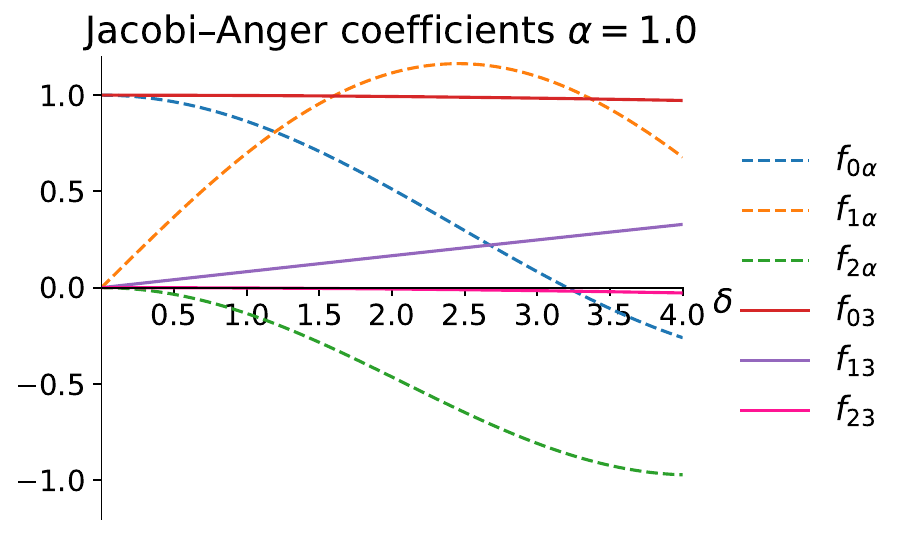}
		    \caption{Magnitude of the different Bessel-function components (without the $\cos$-time-dependence) for the SNAIL depending on the drive amplitude $\delta$ for $\alpha=1.0$. }
            \label{fig:bessel2}
		\end{center}
\end{figure}

For our choice of parameters, all terms with $n>1$ are negligible. Skipping these higher order terms and expanding the potential up to fourth order in the phase $\varphi_c$, we arrive at the SNAIL Hamiltonian,

\begin{align}
\begin{aligned}
    & \mathcal{H}_{\text{SNAIL}} =
    4E_{Cc}n_c^2 \\
     &+  E_J \frac{\varphi_c^2}{2}\left(\alpha f_{0\alpha} + \frac{f_{03}}{3}-\left(\alpha f_{2\alpha} +\frac{f_{23}}{3}\right)\cos(2\omega_Dt) \right) \\
     &-E_J \frac{\varphi_c^4}{24}\left(\alpha f_{0\alpha} + \frac{f_{03}}{3^3}-\left(\alpha f_{2\alpha} +\frac{f_{23}}{3^3}\right)\cos(2\omega_Dt) \right)\\
      &+ E_J\varphi_c \left(\alpha f_{1\alpha}+ f_{13}\right)\cos(\omega_D t) \\
      &- E_J\frac{\varphi_c^3 }{6}\left(\alpha f_{1\alpha}+ \frac{f_{13}}{3^2}\right)\cos(\omega_D t) .
      \label{besselfctnf0f1fsin}
      \end{aligned}
\end{align}  

where $f_{0\alpha} =  J_0(m_l\delta)$, $f_{1\alpha} =  2J_1(m_l\delta)$, $f_{2\alpha} =  2J_2(m_l\delta)$, $f_{03} =  J_0\left(\frac{m_r}{3}\delta\right)$, $f_{13} =  2J_1\left(\frac{m_r}{3}\delta\right)$ and $f_{23} =  2J_2\left(\frac{m_r}{3}\delta\right)$. These coefficients are plotted over the drive amplitudes $\delta$ for $\alpha=0.3$ in Fig. \ref{fig:bessel1}  and for $\alpha = 1.0$ in Fig. \ref{fig:bessel2} for.

As these plots show, the coefficients of the even- and odd-powers of $\varphi$ can be controlled by adjusting the drive amplitude $\delta$.

By choosing $\delta > 0 $ one can enhance the coefficients of the time dependent contributions in Eq. (\ref{besselfctnf0f1fsin}) and thus, the different drive terms. Depending on the value of $\alpha$ that factors into $m_l$ and $m_r$, the Bessel-coefficients are varying differently with the amplitude $\delta$. Hence depending on $\alpha$ the amplitude can be used to enhance or suppress the terms of even or odd power in $\varphi_c$, see Appendix \ref{sec:appendix_driven} for further discussions.

\begin{figure}
		\begin{center}
			\includegraphics[width = 0.45\textwidth]{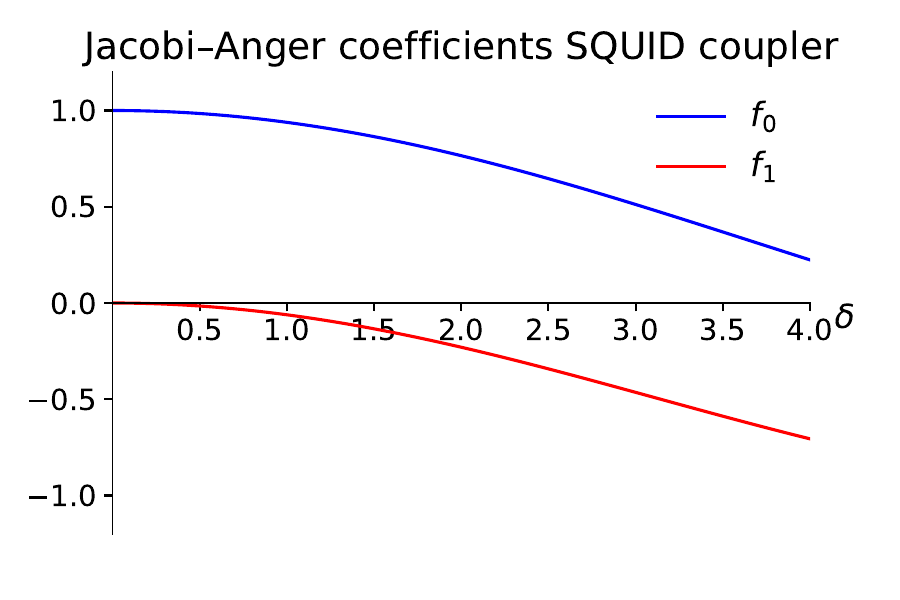}
		    \caption{Magnitude of the different Bessel-function components (without the $\cos$-time-dependence) for the SQUID depending on the drive amplitude $\delta$.}
            \label{fig:bessel_squid} 
		\end{center}
\end{figure}

The Jacobi-Anger expansion can also be done for the dcSQUID. However, due to the symmetry of the junctions, the trigonometric expansion only contains cosine-products, see Eq. (\ref{eq:SQUID_Ham_full}). Therefore, only even power terms of $\varphi_c$ are present and for $n=1$ we get,
\begin{align}
\begin{aligned}
    \mathcal{H}_{\text{dcSQUID}} &=
    4E_{Cc}n_c^2 
    + E_J \varphi_c^2(f_0 - f_2\cos(2\omega_Dt)) \\
    &-E_J\frac{\varphi_c^4}{12}(f_0 - f_2\cos(2\omega_Dt)),
      \label{besselfctndcSQUID}
      \end{aligned}
\end{align}
with $f_0 =J_0\left(\frac{1}{2}\delta\right)$ and $f_2 = 2 J_{2}\left(\frac{1}{2}\delta\right)$. These coefficients can be seen in Fig. \ref{fig:bessel_squid}

The type of gate and the specific qubits involved can be selected by tuning the drive frequency $\omega_D$. In our scheme, we operate with mutually detuned qubits such that, by choosing an appropriate drive frequency, the interaction can be selectively activated. 

Examining the dependence of the prefactors of powers of $\varphi_c$ on the asymmetry factor $\alpha$ and the time-dependent drive parameters, provides insight into how the device parameters and flux pulse must be tuned to implement different types of gates. 

For example, by biasing at a point that enhances the quadratic term, while suppressing the cubic term and driving with a frequency that is, due to the factor 2 in the $\cos$ in Eqs. (\ref{besselfctnf0f1fsin}) and (\ref{besselfctndcSQUID}), half the difference-frequency of two qubits,  $(\omega_i-\omega_j)/2$,  we can get an iSWAP interaction. However, the optimal drive frequency for hitting the resonance, can slightly deviate from this resonance condition, due to the AC Stark shift of the levels $\propto \frac{g^2(\delta)}{\Delta}$ with the detuning of the qubits $\Delta$ and the coupling $g$, which itself depends on the drive amplitude \cite{PhysRevLett.94.123602}.
In a similar manner, one can also determine the parameters for, e.g. multi-qubit interactions.

\section{\label{sec:results} Results}

For both couplers, we simulated the generated dynamics for the models described in Sec.  \ref{sec:theory}, using the parameter sets listed in tables \ref{table:table_squid} and \ref{table:table_snail}.
The simulations were performed using the Python-based quantum simulation package QuTiP \cite{lambert2024qutip5quantumtoolbox}, with basic pulse shaping and minimal optimization. Further improvements in gate fidelity could be achieved using optimal control methods (e.g. GRAPE or CRAB \cite{wilhelm2020introductionoptimalcontrolquantum}) or by incorporating DRAG pulse correction.

The applied pulse envelope was chosen to be a super-Gaussian, which closely resembles the experimentally generated step-like pulses and is given by
\begin{align}
\delta = A \cdot \left(\exp\left[-2^{2r-1}\cdot \log(2) \cdot \left(\frac{(t-t_c)^2}{\sigma^2}\right)^r\right]\right)^2,
\label{eq:supergaussian}
\end{align}
where $A$ is the pulse amplitude, $t_c$ is the pulse center, $\sigma^2$ is the width parameter and $r$ the rank (order) of the Gaussian. 

As described in section \ref{sec:theory}, the gates are implemented via a parametric microwave drive applied to the inductive loop of the respective coupler. The drive frequency required to activate iSWAP-type interaction was derived in Section  \ref{sec:Jacobi-Anger}. Specifically, two-qubit and four-qubit gates can be implemented by driving half the frequency difference between the participating qubits. While in the case of the SNAIL coupler, three-qubit interactions can be driven using the full difference frequency.
All qubits in the simulation are deliberately detuned from each other, to achieve this type of gate scheme. This frequency separation ensures spectral selectivity, allowing the desired interaction to be addressed by applying a chosen drive frequency.

\subsection{Enhanced connectivity: Two qubit gates \label{sec:2qgates}}
\begin{figure}[h]
		\begin{center}
			\includegraphics[width = 0.3\textwidth]{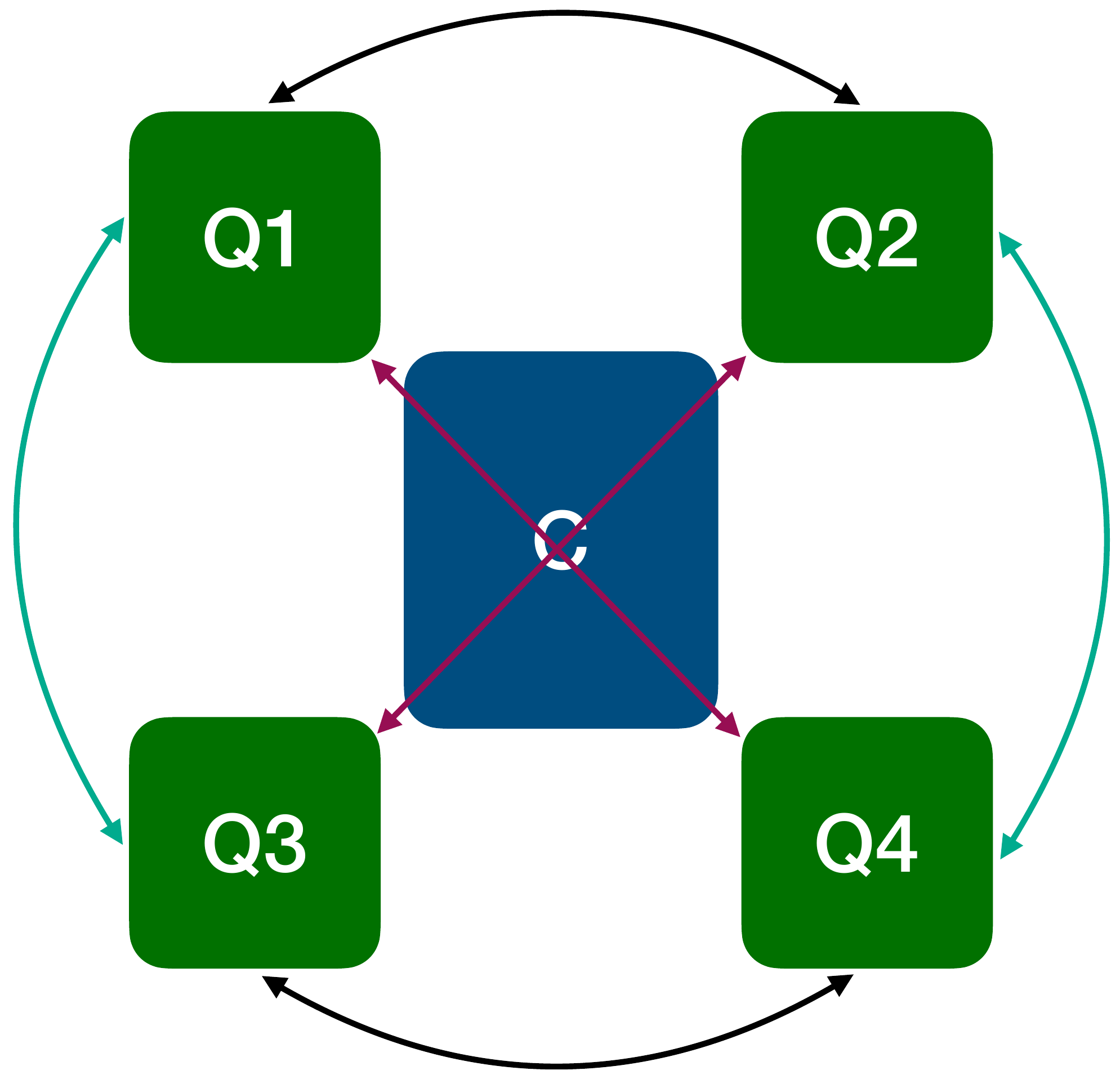}
		    \caption{Via a multi-qubit coupler one can couple pairs of qubits in a horizontal (black), vertical (turquoise) and diagonal (magenta) way by choosing the detunings of the pairs to be different.}
            \label{fig:gate-possibilities}
		\end{center}
		
	\end{figure}

One of the big advantages of multi-qubit couplers, is their all-to-all connectivity within one plaquette, which reduces the number of gates required to do interactions between non-neighboring qubits.
As derived in section \ref{sec:theory}, two-qubit iSWAP-type interactions can be activated by driving at half the frequency difference between the participating qubits. We demonstrate that a multi-qubit coupler enables coupling between qubits that are arranged according to the sketch in Fig. \ref{fig:gate-possibilities} horizontally (qubit 1 and qubit 2, qubit 3 and qubit 4), vertically (qubit 1 and qubit 3, qubit 2 and qubit 4), or diagonally (qubit 1 and qubit 4, qubit 2 and qubit 3). This is possible because the qubits are detuned from one another, allowing the desired pair to be selectively addressed by choosing the corresponding drive frequency.
We focus on an iSWAP-type interaction, for which one needs to drive with the difference of the qubit frequencies, i.e. for
\begin{align}
    \text{interaction:}\quad \hat{a}_i^\dagger\hat{a}_j \quad \leftrightarrow \quad \omega_D = (\omega_i-\omega_j)/2.
    \label{eq:drive_freq_2q_gate}
\end{align}

To evaluate the performance of the system, we compute the gate fidelity. In a multi-qubit setting, it is essential to consider all qubits in the device. Since a gate acting between qubits 1 and 2, for example, must operate correctly regardless of the quantum state of the remaining qubits, i. e. whether qubit 3 or 4 or both are in $|0\rangle$ or $|1\rangle$. In the case of a four-qubit device with only one shared coupler, qubit states can hybridize via the coupler even if they are not directly involved in the gate. Therefore, the fidelity-analysis must be performed on the full four-qubit Hilbert space. The dependence of the targeted transition frequencies on the states of spectator qubits and mutual, residual ZZ interactions between qubits are discussed in more depth in Appendix \ref{sec:alpha_spread}.

The fidelity of an approximation to a unitary in a general $d$-dimensional system can be defined as \cite{baker_single_2022}
\begin{align}
\mathcal{F} = \frac{|\text{Tr}(U_{\text{perfect}}^\dagger U_{\text{simulated}})|}{d},
\label{eq:fidelitymeas}
\end{align}
where $U_{\text{perfect}}$ is the target unitary, $U_{\text{simulated}}$ is the simulated unitary for the time-evolution generated by $H$ and $d$ is the dimension of the systems qubit Hilbert space ($d=16$ for four qubits). To obtain $U_{\text{simulated}}$ we simulate the time evolution in a Hilbert space that takes into account more than the two lowest energy levels per qubit and coupler and then reduce the computed unitary to the computational subspace of the qubits. We checked this procedure for convergence in the subspace truncation.

The ideal target unitary $U_{\text{perfect}}$ is constructed by embedding the desired two-qubit gate into the four-qubit Hilbert space. For example, when implementing a $\sqrt{\text{iSWAP}}$ gate between qubits 1 and 2, the ideal unitary is given by
\begin{align*}
    U_{\text{perfect}} = \sqrt{\text{iSWAP}}_{(1,2)} \otimes I_{(3,4)},
\end{align*}
where $I_{(3,4)}$ is the identity operation on the spectator qubits 3 and 4. And the $\sqrt{\text{iSWAP}}$-gate is given by
\begin{align*}
\sqrt{\text{iSWAP}} =   \begin{pmatrix}
1 & 0 & 0 & 0\\
0 & \frac{1}{\sqrt{2}} & \frac{i}{\sqrt{2}} & 0\\
0 & \frac{i}{\sqrt{2}} & \frac{1}{\sqrt{2}} & 0 \\
0 & 0 & 0 & 1
\end{pmatrix}	
\end{align*}
For comparison with common benchmarks, such as a system with only two qubits or experimental results from two-qubit randomized benchmarking, we also evaluate the fidelity of the active two-qubit subsystem. The subsystem is obtained by tracing out the two spectator qubits and the coupler from the full simulated system, and then computing the fidelity as in Eq. (\ref{eq:fidelitymeas}) for the reduced two-qubit subsystem ($d=4$).  

 We optimize these fidelities by optimizing the parameters of the pulse shape and then correcting the simulated unitary by applying Z-phase gates to all qubits. See also Appendix \ref{sec:full_nm-lagrangian} for an analysis of the generated effective interactions from a normal mode representation of the linearized model.

\subsubsection{ $\sqrt{\text{iSWAP}}$ gates via the dc SQUID}
\begin{figure}[h]
		\begin{center}
			\includegraphics[width = 0.5\textwidth]{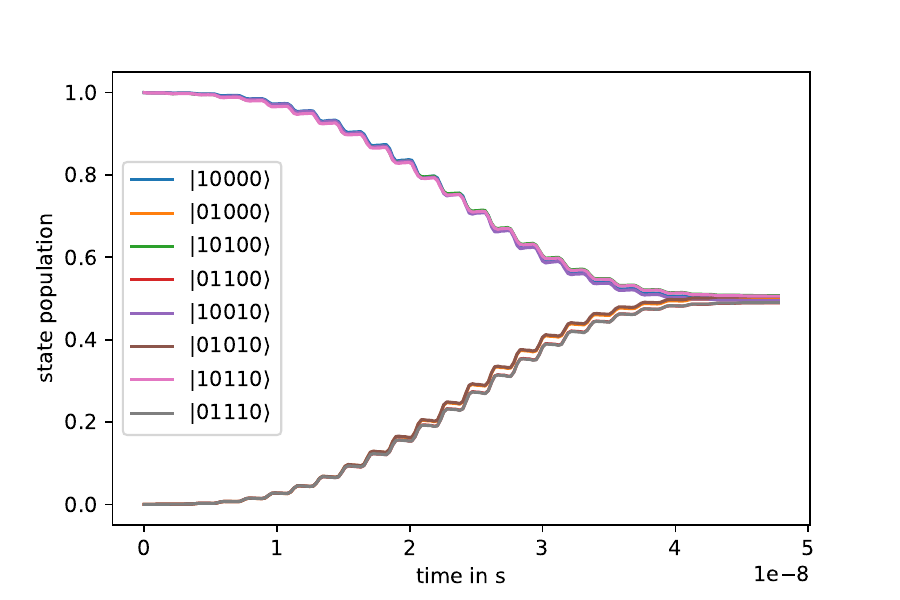}
		    \caption{State populations $|\langle \psi (t)|q_1,q_2,q_3,q_4,c\rangle|^2$ during a $\sqrt{\text{iSWAP}}$-gate between Qubits 1 and 2, activated via the dc-SQUID. The gate duration is 48 ns. We plot the evolution for all configurations of the spectator qubits, see legend, where the notation is $|q_1,q_2,q_3,q_4,c\rangle$.
            The 4-qubit fidelity is 0.9989 and the 2-qubit fidelity with spectator qubits traced-out is 0.9998.The amplitude for the super-Gaussian envelope in equation \ref{eq:supergaussian} is $A = 1.37$.}
            \label{fig:squid12}
		\end{center}
		
	\end{figure}

\begin{figure}[h]
		\begin{center}
			\includegraphics[width = 0.5\textwidth]{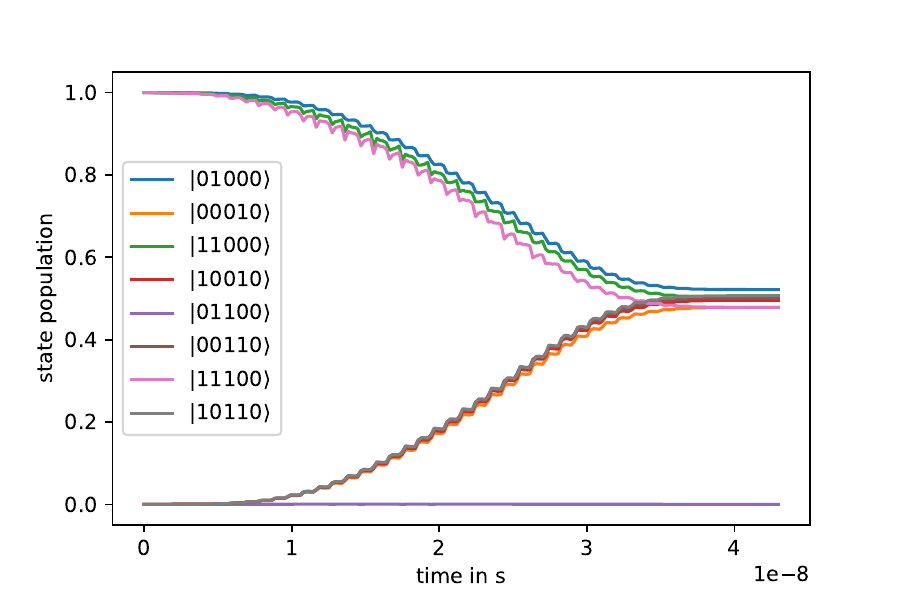}
            \caption{state populations $|\langle \psi (t)|q_1,q_2,q_3,q_4,c\rangle|^2$ during a $\sqrt{\text{iSWAP}}$-gate between Qubits 2 and 4, activated via the dc-SQUID. The gate duration is 38 ns. We plot the evolution for all configurations of the spectator qubits, see legend, where the notation is $|q_1,q_2,q_3,q_4,c\rangle$.
            The 4-qubit fidelity is 0.9950 and the 2-qubit fidelity with spectator qubits traced-out is 0.9988. The amplitude for the super-Gaussian envelope in equation \ref{eq:supergaussian} is $A = 1.53$. }
            \label{fig:squid24}
		\end{center}
	\end{figure}

\begin{figure}[h]
		\begin{center}
			\includegraphics[width = 0.5\textwidth]{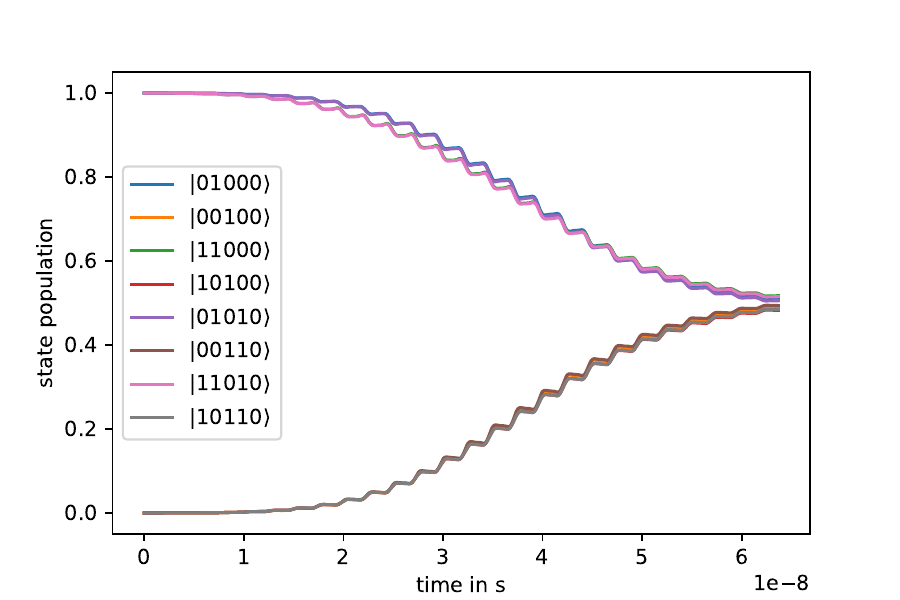}
            \caption{ 
            State populations $|\langle \psi (t)|q_1,q_2,q_3,q_4,c\rangle|^2$ during a $\sqrt{\text{iSWAP}}$-gate between Qubits 2 and 3, activated via the dc-SQUID. The gate duration is 61 ns. We plot the evolution for all configurations of the spectator qubits, see legend, where the notation is $|q_1,q_2,q_3,q_4,c\rangle$.
            The 4-qubit fidelity is 0.9987 and the 2-qubit fidelity with spectator qubits traced-out is 0.9993.The amplitude for the super-Gaussian envelope in equation \ref{eq:supergaussian} is $A = 1.19$.}
            \label{fig:squid23}
		\end{center}
		
	\end{figure}

We simulate different interactions between pairs of qubits and demonstrate that it is possible to couple qubits that are placed horizontally, vertically and diagonally by driving the dc-SQUID with an appropriate frequency, see Eq. (\ref{eq:drive_freq_2q_gate}).
We choose $\varphi_{\text{dc}}=0.$ For labeling the states, we choose the order $|q_1,q_2,q_3,q_4,c\rangle$, where the $q_i$ ($i=1,2,3,4$) label the states of the four qubits and $c$ the states of the coupler.
We show a horizontally placed gate between qubit 1 and qubit 2, by driving with frequency $\omega_D =(\omega_1 -\omega_2)/2$ in Fig. \ref{fig:squid12}. 
As an example for the vertical gate, we show a gate between qubit 2 and qubit 4 by driving with a frequency of $\omega_D =(\omega_2 -\omega_4)/2$ in Fig. \ref{fig:squid24}. 
The diagonal gate can be between qubits 1 and 4 or qubits 2 and 3 and we show the latter here by driving with frequency $\omega_D =(\omega_2 -\omega_3)/2$, see Fig. \ref{fig:squid23}. 

\subsubsection{ $\sqrt{\text{iSWAP}}$  gates via the tunable SNAIL}

In appendices \ref{sec:alpha_spread} and \ref{sec:appendix_driven} we analyze that the standard configuration of e.g.  $\alpha \approx 0.3$ is not ideal to achieve a strong prefactor for the quadratic term of the SNAIL coupler while also keeping the prefactors of the cubic and quartic terms small. When choosing $\varphi_{\text{dc}}=0$, we rather find that the asymmetry parameter should be around $\alpha =1$. Therefore, we use $\alpha= 1$ in the simulations of the two-qubit gates via the SNAIL coupler.
For activating iSWAP-type interactions, one then needs to drive at half the difference of the transition frequencies of the participating qubits, for this configuration, just as in the case of the dc-SQUID coupler, see Eq. \ref{eq:drive_freq_2q_gate}.
We show that it is possible to achieve two-qubit gates in each direction of the qubit layout.

\begin{figure}[h]
		\begin{center}
			\includegraphics[width = 0.5\textwidth]{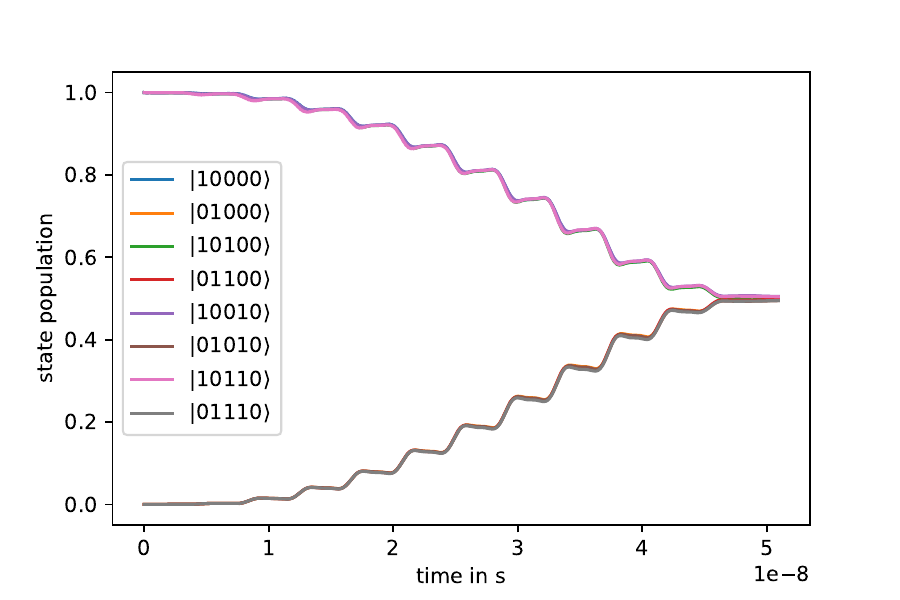}
		   \caption{  
         State populations $|\langle \psi (t)|q_1,q_2,q_3,q_4,c\rangle|^2$ during a $\sqrt{\text{iSWAP}}$-gate between Qubits 1 and 2, activated via the SNAIL. The gate duration is 49 ns. We plot the evolution for all configurations of the spectator qubits, see legend, where the notation is $|q_1,q_2,q_3,q_4,c\rangle$.
            The 4-qubit fidelity is 0.9997 and the 2-qubit fidelity with spectator qubits traced-out is 0.9998.  The amplitude for the super-Gaussian envelope in equation \ref{eq:supergaussian} is $A = 2.46$.}
           	\label{fig:snail12}
		\end{center}
	
	\end{figure}

A horizontally placed gate between qubit 1 and qubit 2 can be achieved by driving with frequency $\omega_D =(\omega_1 -\omega_2)/2$, as shown in Fig. \ref{fig:snail12}. 

\begin{figure}[h]
		\begin{center}
			\includegraphics[width = 0.5\textwidth]{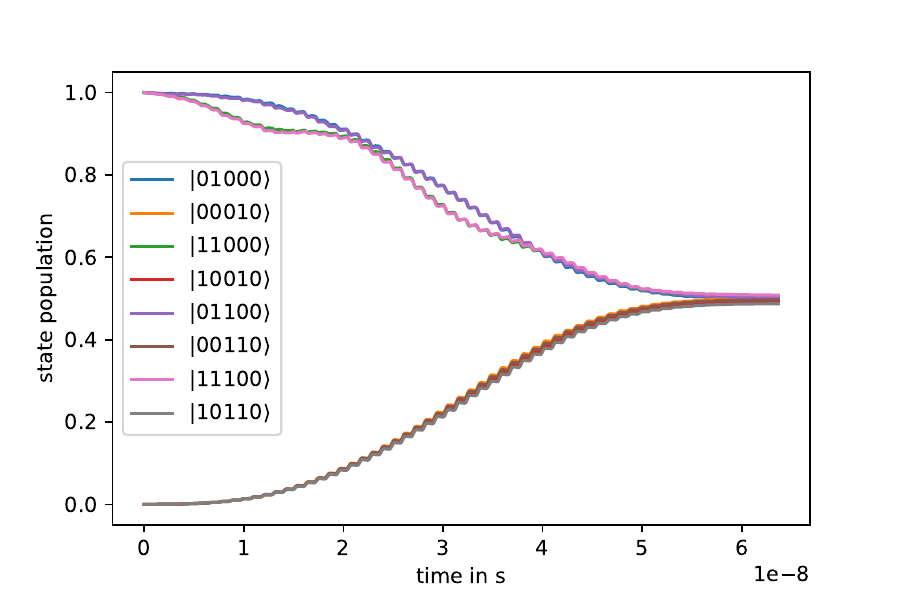}
         \caption{  
         State populations $|\langle \psi (t)|q_1,q_2,q_3,q_4,c\rangle|^2$ during a $\sqrt{\text{iSWAP}}$-gate between Qubits 2 and 4, activated via the SNAIL. The gate duration is 58 ns. We plot the evolution for all configurations of the spectator qubits, see legend, where the notation is $|q_1,q_2,q_3,q_4,c\rangle$.
            The 4-qubit fidelity is 0.9981 and the 2-qubit fidelity with spectator qubits traced-out is 0.9996. The amplitude for the super-Gaussian envelope in equation \ref{eq:supergaussian} is $A = 2.62$.}
            	\label{fig:snail24}
		\end{center}
	
	\end{figure}

\begin{figure}[h]
		\begin{center}
			\includegraphics[width = 0.5\textwidth]{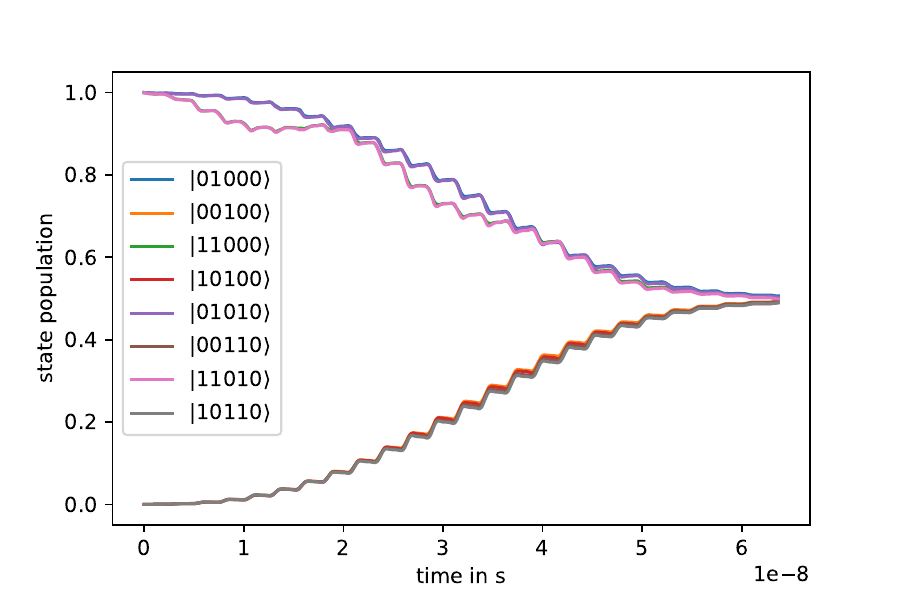}
		    \caption{  
         State populations $|\langle \psi (t)|q_1,q_2,q_3,q_4,c\rangle|^2$ during a $\sqrt{\text{iSWAP}}$-gate between Qubits 2 and 3, activated via the SNAIL. The gate duration is 64 ns. We plot the evolution for all configurations of the spectator qubits, see legend, where the notation is $|q_1,q_2,q_3,q_4,c\rangle$.
            The 4-qubit fidelity is 0.9975 and the 2-qubit fidelity with spectator qubits traced-out is 0.9998. The amplitude for the super-Gaussian envelope in equation \ref{eq:supergaussian} is $A = 2.42$.}
            \label{fig:snail23}
		\end{center}
		
	\end{figure}

As an example for the vertical gate, we show a gate between qubit 2 and qubit 4 by driving with a frequency of $\omega_D =(\omega_2 -\omega_4)/2$, see Fig. \ref{fig:snail24}. 
The diagonal gate can be between qubits 1 and 4 or qubits 2 and 3 and we show the latter here by driving with frequency $\omega_D =(\omega_2 -\omega_3)/2$, see Fig. \ref{fig:snail23}. 

\subsection{Multi-qubit gates}
With the layout used here, it is also possible to implement native multi-qubit gates. As derived in section \ref{sec:theory}, such gates can be realized by driving specific interaction terms in the Hamiltonian with appropriately chosen drive frequencies and amplitudes. In the case of the SNAIL coupler, the interaction can be tailored further by selecting suitable values for the asymmetry parameter $\alpha$ and the applied external dc flux bias $\varphi_{\text{dc}}$. This enables the direct realization of three- and four-qubit interactions without decomposing them into sequences of two-qubit gates, thereby reducing circuit depth and potentially improving overall gate fidelity. 

\subsubsection{Three qubit gate}
In the SNAIL, we can also activate cubic terms in the Hamiltonian, due to the junction asymmetry. In Figs. \ref{fig:spread_all} and \ref{fig:spread_alpha_delta_3q}, we can see that the standard configuration of e.g.  $\alpha \approx 0.3$ is not the perfect configuration for a strong cubic term. Therefore, we choose $\alpha= 0.82$ in the simulation, to minimize the spread but also get a higher coupling strength.
Moreover, as seen in section \ref{sec:Jacobi-Anger}, we need to drive odd terms in $\varphi$ in the Hamiltonian with the full drive frequency.
As an example we apply the drive  $\omega_D = \omega_1-\omega_2-\omega_3$ to achieve the native interaction $\hat{a}_1\hat{a}_2^\dagger\hat{a}_3^\dagger + \text{h.c.}$. In appendix \ref{sec:appendix_driven}, one can see now, that one should choose $\alpha$ to be large and $\varphi_{dc}=0$ for this interaction.

\begin{figure}[h]
		\begin{center}
			\includegraphics[width = 0.5\textwidth]{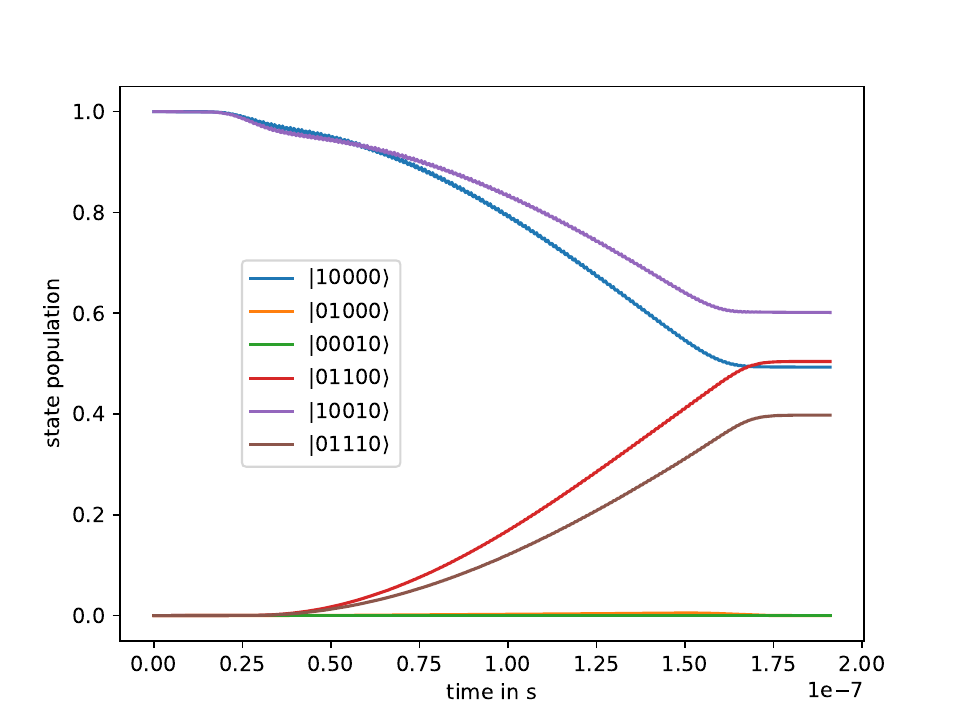}
		    \caption{Native three-qubit gate for the interaction term $\hat{a}_1\hat{a}_2^\dagger\hat{a}_3^\dagger + \text{h.c.}$. We plot the state population  over the time. The gate duration is 200 ns. 
            We plot the evolution for all configurations of the spectator qubits, see legend, where the notation is $|q_1,q_2,q_3,q_4,c\rangle$. 
            The amplitude for the super-Gaussian envelope in equation \ref{eq:supergaussian} is $A = 2.72$. }
            	\label{fig:3q}
		\end{center}
	
	\end{figure}
    
One important quantum simulation application of three-qubit gates is that they can be used for simulating lattice gauge theories (LGTs). 
In the quantum link formulation, as for example in the (2+1)D Abelian Higgs model, the infinite-dimensional gauge field is truncated to a finite-dimensional spin system while local symmetry is preserved \cite{osborne}.
Via a Jordan-Wigner transformation the fermionic matter is mapped to spins, so the gauge-matter coupling is described by a three-body hopping term. The matter fields are modeled as matter qubits at site $n$ and $n+1$ and a gauge field as the link between the qubits, leading to the Hamiltonian,
\begin{align*}
    \hat{H} = \frac{\mu}{2} \sum_{n=1}^L (-1)^n \hat{\sigma}^z_n - J \sum_{n=1}^{L-1} (\hat{\sigma}^+_n\hat{\tau}^+_{n,n+1}\hat{\sigma}^-_{n+1}+h.c.),
\end{align*}
with the mass of the matter sites $\mu$ \cite{busnaina2025nativethreebodyinteractionssuperconducting}.
This correlated hopping term can be modeled by a three-qubit interaction, as we can generate via the SNAIL, since it corresponds to a transition $|1_{m,1}, 0_G, 0_{m,2}\rangle \: \leftrightarrow \:|0_{m,1}, 1_G, 1_{m,2}\rangle$. The interaction here can be modeled by a native three-qubit term, $\hat{a}_1\hat{a}_2^\dagger\hat{a}_3^\dagger + \text{h.c.}$, where the SNAIL is driven by a frequency $\omega_D = \omega_1-\omega_2-\omega_3$ and the matter fields correspond to qubit 1 and 3 and the gauge field  correspond to qubit 2.

For our gate, we get a fidelity of 0.9548 for the full 16x16 matrix.
A decomposition into standard gates of this process via Qiskit \cite{qiskit} yields 54 single qubit and 39 two-qubit gates (CNOT) and 17 p-gates. These standard gates would thus need to have decent fidelities to achieve a comparable performance for the entire sequence, which would however would be significantly slower than our three-qubit gate.

\subsubsection{Four qubit gate}
\begin{figure}[h]
		\begin{center}
			\includegraphics[width = 0.5\textwidth]{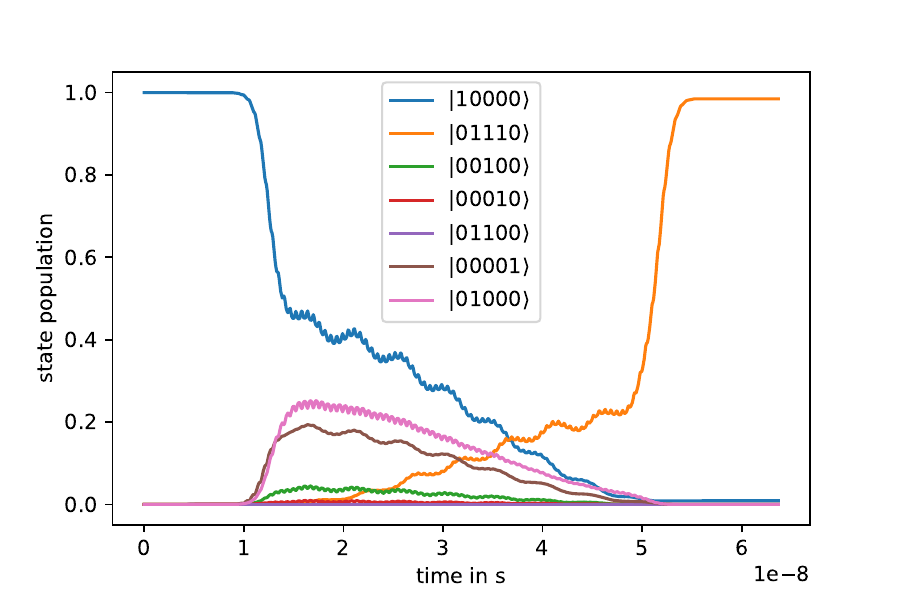}
		    \caption{Native four-qubit gate for the interaction term $\hat{a}_1\hat{a}_2^\dagger\hat{a}_3^\dagger\hat{a}_4^\dagger + \text{h.c.}$. We plot the state population over the time. The gate duration is 46 ns. The amplitude for the super-Gaussian envelope in equation \ref{eq:supergaussian} is $A =  2.95$ }
            \label{fig:4q}
		\end{center}
		
	\end{figure}

In the SQUID, as well as the SNAIL, we can also generate interactions that involve all four qubits. Since these interactions are quite weak, the drive frequency needs to be chosen carefully. For the SNAIL, one also has to choose the parameters carefully, since the quadratic and cubic terms have stronger prefactors in the Hamiltonian. 
We show in Fig. \ref{fig:4q} one example of a native four qubit gate, by driving the interaction $\hat{a}_1\hat{a}_2^\dagger\hat{a}_3^\dagger\hat{a}_4^\dagger + \text{h.c.}$ in the Hamiltonian of the SQUID. To achieve this, we need to drive with the frequency $\omega_D = (\omega_1-\omega_2-\omega_3-\omega_4)/2 \approx (\omega_{|10000\rangle}-\omega_{|01110\rangle})/2$, where the notation for the states is denoted as $|q_1,q_2,q_3,q_4,c\rangle$ and the corresponding frequency of the state is $\omega_{|q_1,q_2,q_3,q_4,c\rangle}$.

A decomposition of this gate into standard gates via Qiskit \cite{qiskit} yields 131 single qubit and 76 two-qubit gates (CNOT). Here, we see a fidelity of 0.7930, which could still be improved with pulse optimization and optimal parameter tuning.

\subsection{Simultaneous execution of two $\sqrt{\text{iSWAP}}$ gates}
We also test, whether it is possible to execute two gates simultaneously. To do so, we choose here the example of executing two horizontal gates simultaneously. Yet other cases would also be possible.
The ideal unitary for these gates is,
\begin{align*}
    U_{\text{perfect}} = \sqrt{\text{iSWAP}}_{(1,2)} \otimes \sqrt{\text{iSWAP}}_{(3,4)}.
\end{align*}
For executing these gates, we use a two tone pulse with two amplitudes, that can overlap.
The two drive frequencies are
$\omega_{D_1} = (\omega_1-\omega_2)/2$ and $\omega_{D_2} = (\omega_3-\omega_4)/2$.
The fidelity of the two simultaneously executed $\sqrt{\text{iSWAP}}$-gates is  $\mathcal{F}=0.9935$ for the SQUID-coupler.
The gate fidelity could be further improved with pulse optimization and optimal control.

\begin{figure}[h]
		\begin{center}
			\includegraphics[width = 0.5\textwidth]{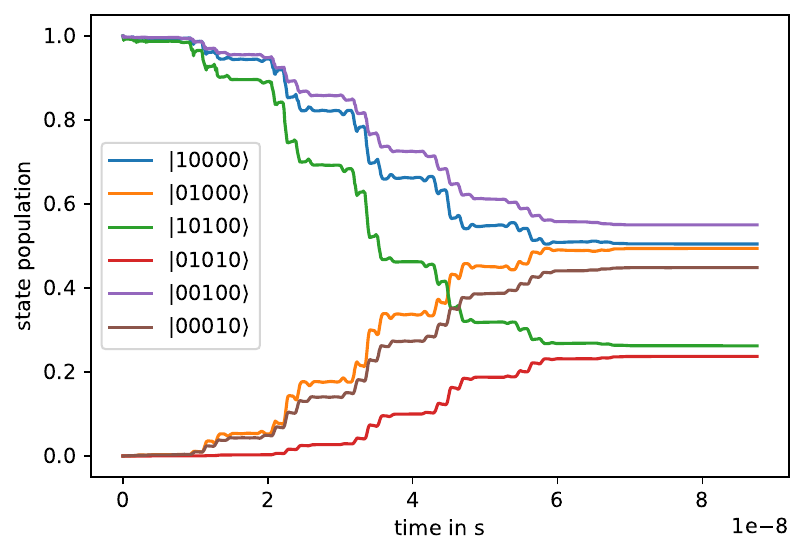}
		    \caption{Simultaneous execution of a $\sqrt{\text{iSWAP}}$-gate between qubit 1 and 2 and a $\sqrt{\text{iSWAP}}$-gate between qubit 3 and 4 via the SQUID. The amplitude for the super-Gaussian envelope in equation \ref{eq:supergaussian} is $A = 0.96$ for the one pulse and $A=0.62$ for the other one. }
            \label{fig:two_tone}
		\end{center}
		
	\end{figure}

\section{\label{sec:problems-errors} Discussion}
Multi-qubit couplers offer several advantages over two-qubit couplers, primarily better connectivity and flexibility.
However, there are still some challenges that could be further improved upon by, for example, optimal control techniques or even alternative designs.
One challenge is that the difference between the energies of the initial and final states of a gate, and therefore the desired frequency of the parametric drive, have a dependence on the states of spectator qubits due to residual cross talk. We for example find that these vary by $\sim 50$kHz for the $\sqrt{\text{iSWAP}}$ gate between qubits 1 and 2. The consequences can for example be seen in the deviation between the different lines in Fig. \ref{fig:squid12}.

Furthermore, since the coupler only has one degree of freedom, suppressing residual ZZ-interactions is non-trivial. Moreover, the detunings must also be chosen such that higher-order frequency detunings are not excited in the very large frequency space.

\section{\label{sec:conclusions} Conclusions}
In summary, we propose two different configurations for implementing a multi-qubit coupler layout, one with an $\alpha$-tunable SNAIL and one with a symmetric dc-SQUID.
We derived the time-dependent Hamiltonians with the help of the irrotational gauge and a Jacobi-Anger expansion. For these, we determine the drive frequencies for activating the different interactions generated by the considered couplers.

We find regimes in the SNAIL parameter space, where different types of gates are favorable, and we thus can also implement multi-qubit gates, such as three or four qubit gates. These can be crucial building blocks for the simulation of lattice gauge theories. 
Due to the parametric driving scheme, we can drive pairwise two-qubit interactions, making the multi-qubit coupler scheme very flexible.
For the $\sqrt{\text{iSWAP}}$-gates in the SNAIL and SQUID cases we find gate times of maximum 64 ns and minimal fidelities  of $0.9950$ for the full 16x16 qubit-subspace. For better comparison to two-qubit couplers and experimental results from two-qubit randomized benchmarking, we obtain the subsystem fidelity by tracing out the two spectator qubits. Here we achieve minimal fidelities of $0.9988$. 
We are also able to drive three and four qubit gates, where the achieved fidelities have slightly lower values. Moreover, it is possible to execute two gates in parallel. Here, however, the fidelities do not match up two single two-qubit gates, meaning there is some cross-talk between both operations. 

Improving upon this finding with either a different gate scheme or better optimization, is thus an interesting task for future research. Our analysis shows certain limitations inherent to the current scheme with only one tunable element to control interactions between four qubits. Future work should therefore investigate alternative schemes, like tunable couplings and tunable qubits or the integration of couplers with more than one tunable degree of freedom, which may offer enhanced controllability, the possibility to suppress or even cancel crosstalk and thus improve overall system performance further.

This work thus shows that multi-qubit couplers can be used as a flexible and versatile tool for simulating reasonably good two-qubit gates, as well as multi-qubit gates. Which significantly reduced the number of couplers needed, as well as the number of gates for non-neighboring gate executions.

\begin{acknowledgments}

This work received support from the German Federal Ministry of Education and Research via the funding program quantum technologies - from basic research to the market under contract number 13N15684 ``GeQCoS". It is also part of the Munich Quantum Valley, which is supported by the Bavarian state government with funds from the Hightech Agenda Bayern Plus.
\end{acknowledgments}

\appendix

\section{Spectator and crosstalk-analysis for the circuit asymmetry parameter $\alpha$
\label{sec:alpha_spread}}
In a multi-qubit setup, it is important that the gates between qubits work irrespective of the state of adjacent spectator qubits. 
Our gates are parametrically driven, consequently the gates work best if the resonance frequency is independent of the states of the spectator qubits. Therefore we analyze the dependence of the transition frequency, that we aim to drive, on the state of all the spectator qubits in the four-qubit coupler system.
The spread of the resonances reads,
\begin{align}
    \Delta \omega = \max_s \omega_s -\min_s \omega_s,
    \label{eq:spread}
\end{align}
where $\omega_s$ is the drive frequency required for resonant  population transfer, for example $\omega_s = (\omega_{|1_1,0_2,s\rangle}-\omega_{|1_1,0_2,s\rangle})/2$ for a gate between q1 and q2,  and $s$ labels the states of the spectator qubits, for a gate between q1 and q2 $s=\{|0_3,0_4\rangle, |1_3,0_4\rangle, |0_3,1_4\rangle,|1_3,1_4\rangle\}$). $\Delta\omega$ is thus an upper bound to the mismatch of the drive frequency to the transition frequency for all spectator configurations. 
To take ac-Stark shifts into account, we consider the time-averaged version of the driven Hamiltonian, which includes the time independent terms from the Jacobi-Anger expansion,
\begin{align*}
    \mathcal{H}_{\text{SNAIL}} =
    4E_{Cc}n_c^2 & +  E_J \frac{\varphi_c^2}{2}\left(\alpha f_{0\alpha}
    + \frac{f_{03}}{3} \right)\\
    & -E_J \frac{\varphi_c^4}{24}\left(\alpha f_{0\alpha} + \frac{f_{03}}{3^3}\right),
\end{align*}
with  $f_{0\alpha} =  J_0(m_l\delta)$, $f_{03} =  J_0\left(\frac{m_r}{3}\delta\right)$, to compute the frequencies $\omega_{|j_1,j_2,j_3,j_4}$. 

\begin{figure}
\centering
        \includegraphics[width = 0.5\textwidth]{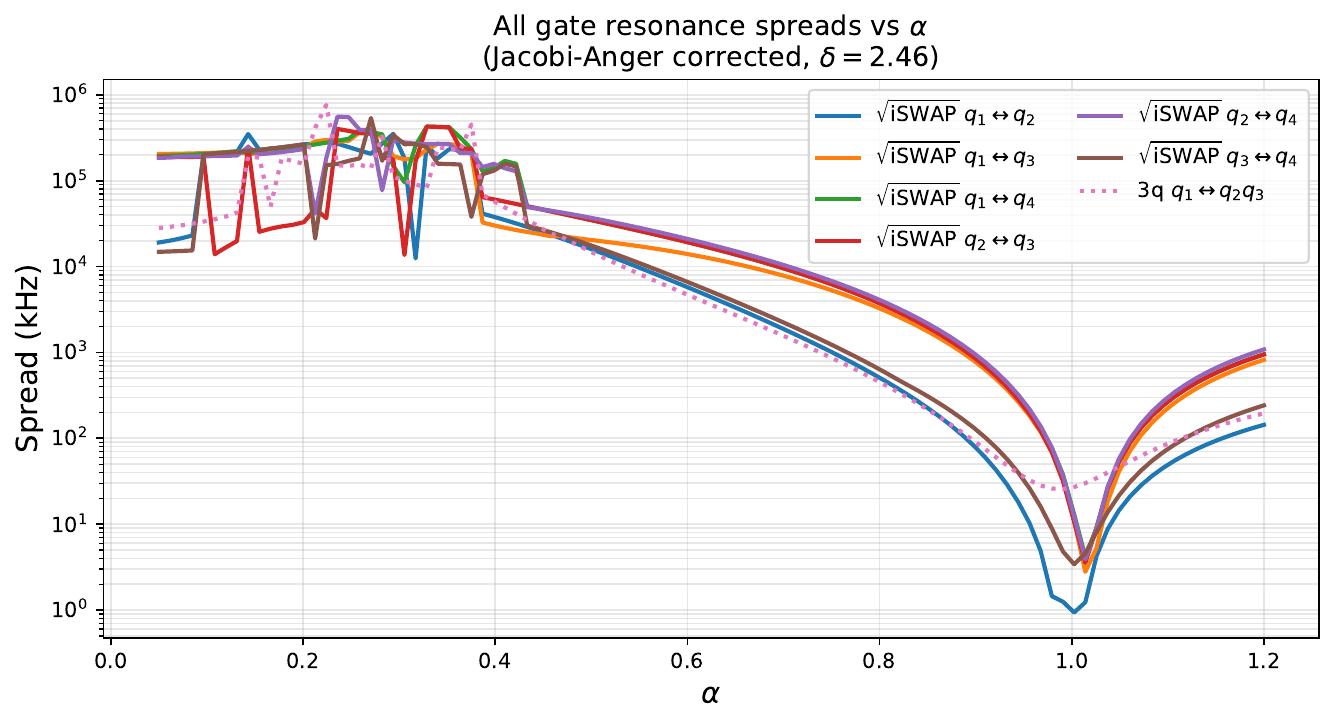}
		    \caption{Resonance spread of all two-qubit gate combinations we can achieve and one three-qubit gate with the Jacobi-Anger drive-corrected frequencies at the drive-strength of the q1 $\leftrightarrow$ q2 gate. The spread is plotted logarithmically in kHz. We can see that in this Jacobi-Anger corrected frequency case, the spread is minimal at $\alpha \approx 1$.
            } 
            \label{fig:spread_all}
\end{figure}

The spread as defined in equation \ref{eq:spread} for the $\sqrt{\text{iSWAP}}$ gates for all possible two-qubit combinations, and the 3-qubit gate we consider, is plotted in Fig. \ref{fig:spread_all}.

\begin{figure}
\centering
        \includegraphics[width = 0.5\textwidth]{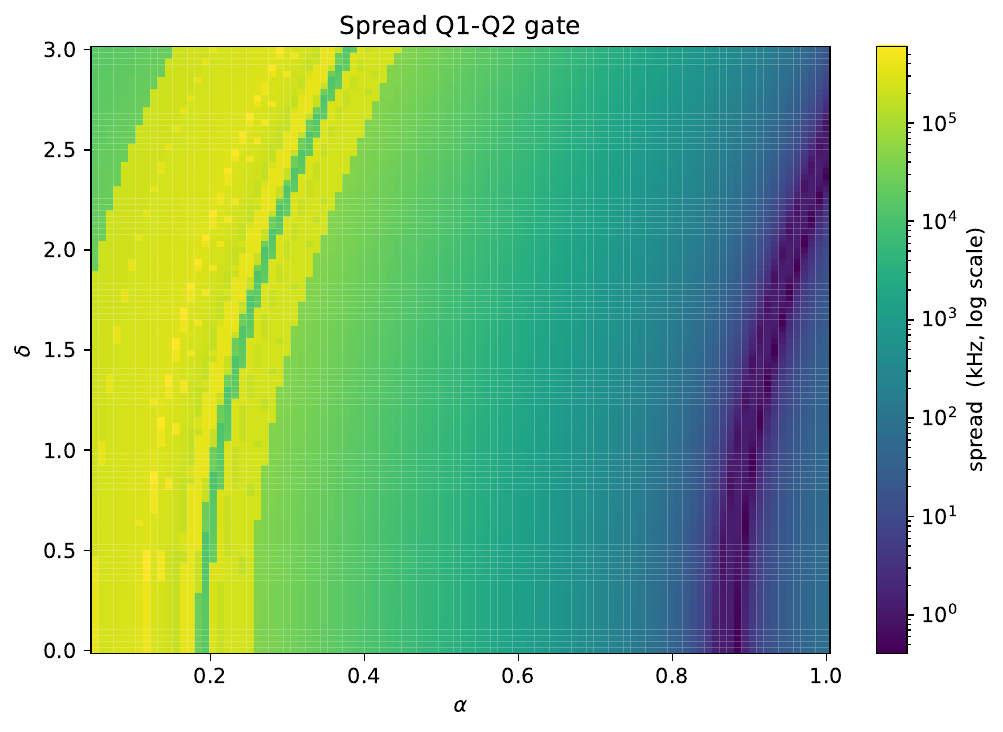}
		    \caption{Resonance spread in kHz for the two-qubit $\sqrt{\text{iSWAP}}$- gate between q1 and q2 over the asymmetry parameter $\alpha$ of the SNAIL and the drive amplitude $\delta$ on a logarithmic scale. }
            \label{fig:spread_alpha_delta_12}
\end{figure}

\begin{figure}
\centering
        \includegraphics[width = 0.5\textwidth]{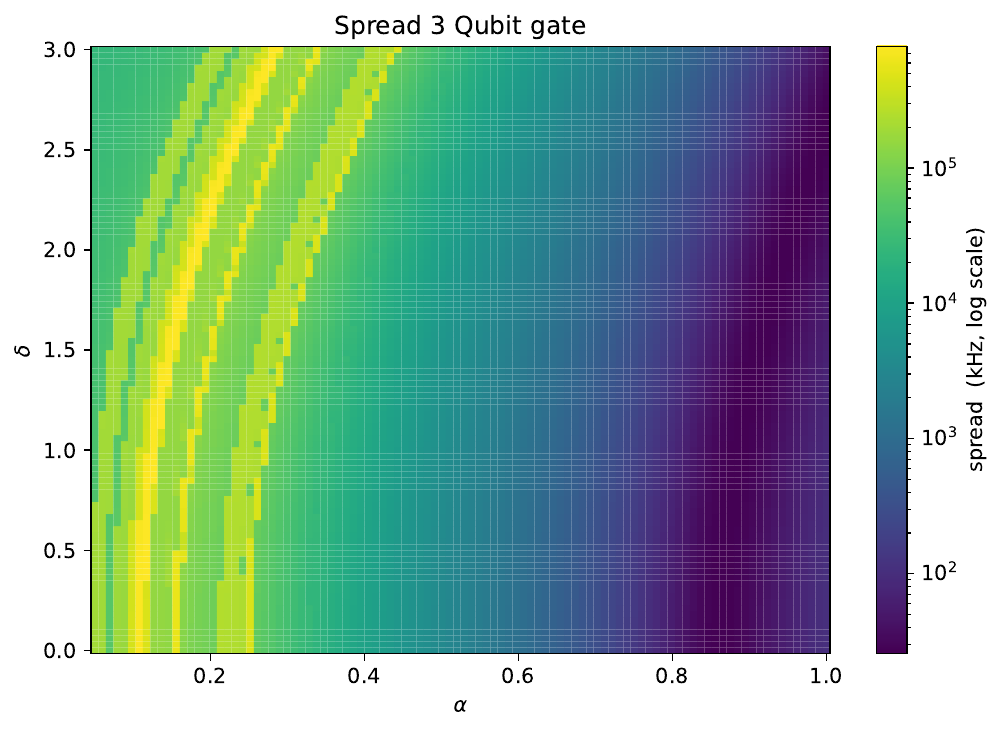}
		    \caption{Resonance spread in kHz for the three-qubit interaction $\hat{a}_1\hat{a}^\dagger_2\hat{a}^\dagger_3 + \text{h. c.}$ over the asymmetry parameter $\alpha$ of the SNAIL and the drive amplitude $\delta$ on a logarithmic scale. }
            \label{fig:spread_alpha_delta_3q}
\end{figure}

Through the Jacobi-Anger coefficients, the spread also depends on the amplitude of the drive. Therefore we plot the dependence of the spread on the  asymmetry parameter $\alpha$ and the drive amplitude $\delta$ in Fig. \ref{fig:spread_alpha_delta_12} for the q1-q2  gate and in Fig. \ref{fig:spread_alpha_delta_3q} for the three-qubit gate.

We can see that the best asymmetry parameters to minimize the spread are for values $\alpha \geq 0.85$. However, we can see that for smaller $\alpha$-values the drive amplitude needs to be smaller, thus the gate-time will be larger.


\begin{figure}[h]
\centering
        \includegraphics[width = 0.5\textwidth]{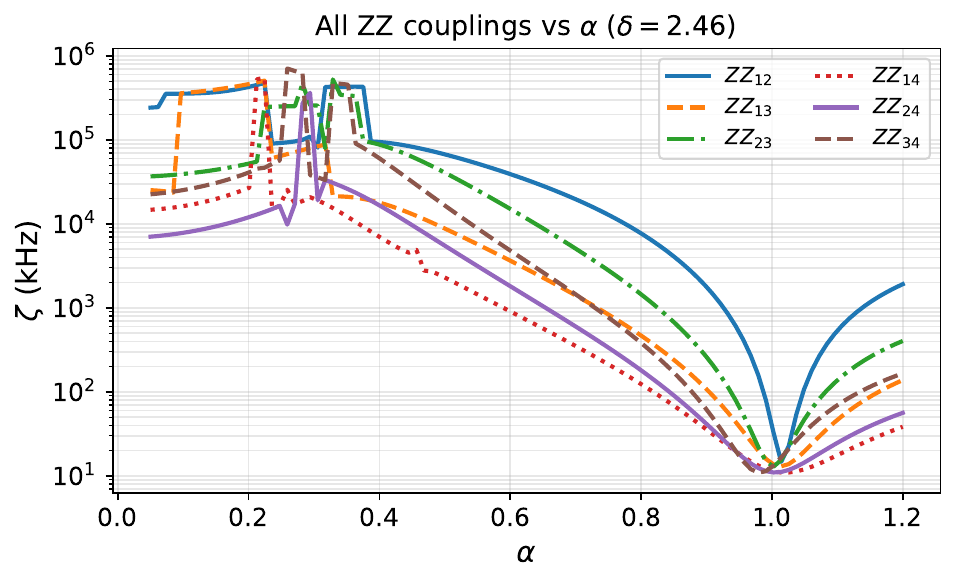}
		    \caption{ZZ-crosstalk $\zeta$ in kHz between all pairs of qubits vs. the circuit asymmetry parameter $\alpha$. The ZZ-crosstalk is corrected for the driven system by the Jacobi-Anger drive-corrected frequencies at the drive-strength of the q1 $\leftrightarrow$ q2 gate. $\zeta$ is minimal at $\alpha \approx 1$ for all possible combinations of two-qubit ZZ-crosstalk. }
            \label{fig:ZZ}
\end{figure}

Since the spread also depends on the two-qubit ZZ-crosstalk, we simulate this too in the static case as a function of the asymmetry parameter $\alpha$.
We calculate the static ZZ from the dressed energy eigenvalues in the QuTiP-simulation for all qubit pairs \cite{PhysRevApplied.15.064074}
\begin{align*}
    \zeta = E_{11}- E_{10}-E_{01} +E_{00},
\end{align*}
In figure \ref{fig:ZZ}, we can see that the static ZZ minimum aligns with the static resonance spread minimum and is the leading cause in the misalignment of resonance frequencies with spectators.
However, in the driven case, other effects like AC Stark shifts contribute.

\section{Normal mode derivation of the SNAIL Lagrangian 
\label{sec:normal_modes}}
We further want to analyze and understand the various interactions that can be generated via a SNAIL coupler in the multi-qubit system. For this, we compute the normal modes of the linearized system Lagrangian. In doing so we also account for the time-averaged effect of the $\varphi_{ac}$ on the SNAIL potential. 

The original Lagrangian is given by,

\begin{align*}
  \mathcal{L} & = \sum_{i=1}^4 \left[\frac{C_i}{2}\dot{\phi}_i^2 +\frac{C}{2}(\dot{\phi}_i -\dot{\phi}_c)^2 + E_{J_i}\cos\left(\frac{2\pi}{\phi_0}\phi_i\right) \right] \\
    & +\frac{C_c}{2}\dot{\phi}_c^2 - U(\phi_c,\phi_{\text{ext}}(t)) , 
\end{align*}
where the SNAIL potential in the irrotational gauge \ref{sec:tdep} is given by,
\begin{align}
& U = -\alpha E_{J}\cos\left(\frac{2\pi}{\phi_0}\left(\phi_c+m_l\phi_{\text{ext}}(t)\right)\right) 
\\&-3 E_J\cos\left(\frac{2\pi}{\phi_0}\left(\frac{\phi_c+m_r\phi_{\text{ext}}(t)}{3}\right)\right).
\end{align}
Expanding the cosine terms via the Jacobi-Anger expansion and keeping only the leading order (analogous to the Hamiltonian \ref{besselfctnf0f1fsin}), we find,
\begin{align}
\begin{aligned}
    & \mathcal{L}_{\text{SNAIL}} =
    \frac{1}{2}C_c \dot{\phi}_c \\
     &-  \frac{E_J}{2} \left(\frac{2\pi\phi_c}{\phi_0}\right)^2\left(\alpha f_{0\alpha} + \frac{f_{03}}{3}-\left(\alpha f_{2\alpha} +\frac{f_{23}}{3}\right)\cos(2\omega_Dt) \right) \\
     &+\frac{E_J}{24}\left(\frac{2\pi\phi_c}{\phi_0}\right)^4\left(\alpha f_{0\alpha} + \frac{f_{03}}{3^3}-\left(\alpha f_{2\alpha} +\frac{f_{23}}{3^3}\right)\cos(2\omega_Dt) \right)\\
      &- E_J\left(\frac{2\pi\phi_c}{\phi_0}\right) \left(\alpha f_{1\alpha}+ f_{13}\right)\cos(\omega_D t) \\
      &+ \frac{E_J}{6}\left(\frac{2\pi\phi_c}{\phi_0}\right)^3\left(\alpha f_{1\alpha}+ \frac{f_{13}}{3^2}\right)\cos(\omega_D t) .
      \end{aligned}
\end{align}  
where $f_{0\alpha} =  J_0(m_l\delta)$, $f_{1\alpha} =  2J_1(m_l\delta)$, $f_{2\alpha} =  2J_2(m_l\delta)$, $f_{03} =  J_0\left(\frac{m_r}{3}\delta\right)$, $f_{13} =  2J_1\left(\frac{m_r}{3}\delta\right)$ and $f_{23} =  2J_2\left(\frac{m_r}{3}\delta\right)$.

We eliminate the explicit time-dependence by performing an average over one period of the drive, where all oscillatory terms vanish. In this approximation we get
\begin{align*}
    \overline{\mathcal{L}}_{\text{SNAIL}}
    =\frac{1}{2}C_c \dot{\phi}_c &-\frac{c_2^{(0)}}{2} \phi_c^2 + \frac{c_4^{(0)}}{24}\phi^4\\
    \end{align*}
with $c_2^{(0)} = E_J \left(\frac{2\pi}{\phi_0}\right)^2 \left[\alpha f_{0\alpha} +\frac{f_{03}}{3}\right]$ and $c_4^{(0)} = E_J \left(\frac{2\pi}{\phi_0}\right)^4 \left[\alpha f_{0\alpha} +\frac{f_{03}}{3^3}\right]$.

For finding the normal modes, we need to look at the linear system and therefore introduce the abbreviation
\begin{align*}
    E_{Jc}= E_J\left[\frac{1}{3} f_{03}+\alpha f_{0\alpha}\right],
\end{align*}
where at zero drive-amplitude $\delta=0$, one recovers the static result ($J_0(x) =1$).
We further define the inductances as $L_i = \left(\frac{\phi_0}{2\pi}\right)^2 E_{Ji}^{-1}$.

With this, we can write the linear Lagrangian as,
\begin{align*}
    \mathcal{L}_{\text{lin}}=\frac{1}{2}\dot{\vec{\phi}}^T[C]\dot{\vec{\phi}} - \frac{1}{2}\vec{\phi}^T[L^{-1}]\vec{\phi}, 
\end{align*}
where $[C]$ is the capacitance matrix and $[L^{-1}]$ is the inverse inductance matrix.
So we need to solve $[L^{-1}] -\Omega^2[C]=0$ to find its normal modes.
The capacitance matrix and the inductance matrix do not commute, so we can define scaled coordinates $ \Phi_j = \phi_j / \sqrt{L_j}$,
and the linear part of the Lagrangian reads
\begin{align*}
    \mathcal{L}= \frac{1}{2}\dot{\Phi}_iA_{ij}\dot{\Phi}_j - \frac{1}{2}\Phi \delta_{ij}\Phi_j \, ,
\end{align*}
where
\begin{align*}
A_{ij}=    
\begin{pmatrix}
\frac{1}{\Omega_1^2} & 0 & 0 & 0 & \frac{-\beta_{1c}}{\Omega_1\Omega_c}\\
 0 & \frac{1}{\Omega_2^2}  &  0& 0 &  \frac{-\beta_{2c}}{\Omega_2\Omega_c} \\ 
0& 0& \frac{1}{\Omega_3^2}&0 &\frac{-\beta_{3c}}{\Omega_3\Omega_c} \\
0 & 0 & 0 &
\frac{1}{\Omega_4^2}& \frac{-\beta_{4c}}{\Omega_4\Omega_c} \\
\frac{-\beta_{c1}}{\Omega_c\Omega_1} & \frac{-\beta_{c2}}{\Omega_c\Omega_2} & \frac{-\beta_{c3}}{\Omega_c\Omega_3} &
\frac{-\beta_{c4}}{\Omega_c\Omega_4} & 
\frac{1}{\Omega_c^2}
\end{pmatrix}.	
\end{align*}
with $\beta_{ic}=\beta_{ci}=\frac{C}{\sqrt{(C_i+C)(C_c+4C)}}$, $\Omega_i = \frac{1}{\sqrt{L_i(C_i+C)}}$ for $i=1,...,4$ and $\Omega_c = \frac{1}{\sqrt{L_c(C_c+4C)}}$.

We solve for the eigensystem of this matrix, with the help of perturbation theory, making use of the fact, that the coupling capacitances $C$ are way smaller than the capacitances of the transmons and SNAIL. 
The ansatz for the perturbation theory up to first order is,
\begin{align*}
    A = A_0 + C\cdot A_1,
\end{align*}
with 
\begin{align*}
    A_0 = \diag \left(L_1C_1,L_2C_2,L_3C_3,L_4C_4,L_cC_C\right)
\end{align*}
 and
 \begin{widetext}
\begin{align*}
A_1=    
\begin{pmatrix}
L_1 & 0 & 0 & 0 & -\frac{\sqrt{L_1C_1}\sqrt{L_cC_c}}{\sqrt{C_1C_c}}\\
 0 & L_2  &  0& 0 &  -\frac{\sqrt{L_2C_2}\sqrt{L_cC_c}}{\sqrt{C_2C_c}} \\ 
0& 0& L_3&0 &-\frac{\sqrt{L_3C_3}\sqrt{L_cC_c}}{\sqrt{C_3C_c}} \\
0 & 0 & 0 &
L_4& -\frac{\sqrt{L_4C_4}\sqrt{L_cC_c}}{\sqrt{C_4C_c}} \\
-\frac{\sqrt{L_1C_1}\sqrt{L_cC_c}}{\sqrt{C_1C_c}} & -\frac{\sqrt{L_2C_2}\sqrt{L_cC_c}}{\sqrt{C_2C_c}} & -\frac{\sqrt{L_3C_3}\sqrt{L_cC_c}}{\sqrt{C_3C_c}} &
-\frac{\sqrt{L_4C_4}\sqrt{L_cC_c}}{\sqrt{C_4C_c}} & 
4L_c
\end{pmatrix} ,	
\end{align*}
 \end{widetext}
where we needed to do an expansion in $C$ to first order for the $\frac{1}{\Omega_i}\Big|_{i=1,...4,c}$ and  $\frac{-\beta_{ic}}{\Omega_i\Omega_c}\Big|_{i=1,...4}$ terms.
With eigenvectors $|v_i^{(0)}\rangle =\{e_i\}_{i=1}^n$ for the matrix $A_0$ and eigenvalues $\nu_1^{(0)}=L_1C_1,\,\nu_2^{(0)}=L_2C_2,\,\nu_3^{(0)}=L_3C_3,\,\nu_4^{(0)}=L_4C_4,\,\nu_c^{(0)}=L_cC_c$. 
The first order corrections to the eigenvalues are $\nu_i^{(1)}= \langle v_i^{(0)}|A_1|v_i^{(0)}\rangle$.   
Up to first order, we thus get for the eigenvalues
$\nu_i = \nu_i^{(0)} + C\cdot  \nu_i^{(1)} =
\{L_1(C_1+C),L_2(C_2+C),L_3(C_3+C),L_4(C_4+C),L_c(C_c+4C)\}.$

The first order correction to the eigenvectors is 
\begin{align*}
    |v_i^{(1)}\rangle = \sum_{i\neq j} \frac{\langle v_j^{(0)}|A_1|v_i^{(0)}\rangle }{\nu_i^{(0)}-\nu_j^{(0)}}|v_j^{(0)}\rangle ,
\end{align*}
and hence the eigenvectors up to first order, $|v_i\rangle = |v_i^{(0)}\rangle + C\cdot |v_i^{(1)}\rangle$, are,
\begin{align*}
\vec{v}_1 = 
\begin{pmatrix}
    1 \\
    0\\
    0\\
    0\\
    -v_1
\end{pmatrix}, 
    \vec{v}_2 =  
\begin{pmatrix}
    0\\
    1\\
    0\\
    0\\
   -v_2
\end{pmatrix}, 
\vec{v}_3 =  \begin{pmatrix}
    0\\
    0\\
    1\\
    0\\
    -v_3
\end{pmatrix}, \\
\vec{v}_4 = \begin{pmatrix}
    0\\
    0\\
    0\\
    1\\
    -v_4
\end{pmatrix}, 
\vec{v}_c =  
\begin{pmatrix}
     -v_1\\
     -v_2\\ 
     -v_3\\
     -v_4\\
     1
     \end{pmatrix}.
\end{align*}
Where $v_{i} =\frac{C\sqrt{C_iL_i}\sqrt{C_cL_c}}{\sqrt{C_iC_c}(C_iL_i -C_cL_c)}$.

We can normalize these, and define the normalized eigenvectors as $\vec{b}_1, \vec{b}_2,\vec{b}_3,\vec{b}_4,\vec{b}_c$. These can be grouped in a matrix,
\begin{align*}
    B=\left(\vec{b}_1, \vec{b}_2, \vec{b}_3, \vec{b}_4,\vec{b}_c\right) = \left(\begin{array}{ccccc}
        B_{11} & 0&0&0&B_{1c} \\
        0 & B_{22}&0&0&B_{2c} \\
        0 & 0&B_{33}&0&B_{3c} \\
        0 & 0&0&B_{44}&B_{4c} \\
        B_{c1} & B_{c2}&B_{c3}&B_{c4}&B_{cc} 
    \end{array}\right). 
\end{align*}

Therefore, the transformation to normal modes is given by
\begin{align}
    \Phi=B\Psi \leftarrow \text{normal modes}.
    \label{eq:nm_eq}
\end{align}
which can be inserted into the full Lagrangian, that includes the time-dependence.

\section{Full Lagrangian in normal mode representation \label{sec:full_nm-lagrangian}}
Since, we are driving the system with a time-dependent pulse, we now need to plug in the  normal modes from Eq. (\ref{eq:nm_eq}) into the time-dependent Lagrangian. This means we now include the coefficients $f_{13}, f_{1\alpha}, f_{23}$ and $f_{2\alpha}$ in the expansion-coefficients
\begin{align}
    \begin{aligned}  
        &c_1^{(1)}  = E_J \frac{2\pi}{\phi_0}\left[f_{13} +\alpha f_{1\alpha}\right]\\
        &c_2^{(2)}= E_J \left(\frac{2\pi}{\phi_0}\right)^2 \left[\frac{f_{23}}{3} +\alpha f_{2\alpha}\right] \\
        &c_3^{(1)} =  E_J \left(\frac{2\pi}{\phi_0}\right)^3 \left[\frac{f_{13}}{3^2} +\alpha f_{1\alpha}\right]\\
        &c_4^{(2)} = E_J \left(\frac{2\pi}{\phi_0}\right)^4\left[\frac{f_{23}}{3^3} +\alpha f_{2\alpha}\right], 
    \end{aligned}
    \label{eq:coeff_t-dep}
\end{align}
where $f_{0\alpha} =  J_0(m_l\delta)$, $f_{1\alpha} =  2J_1(m_l\delta)$, $f_{2\alpha} =  2J_2(m_l\delta)$, $f_{03} =  J_0\left(\frac{m_r}{3}\delta\right)$, $f_{13} =  2J_1\left(\frac{m_r}{3}\delta\right)$ and $f_{23} =  2J_2\left(\frac{m_r}{3}\delta\right)$.
The time-dependent Lagrangian in terms of normal modes thus reads,
\begin{widetext}
    \begin{align}
\begin{aligned}
  \mathcal{L} = &\sum_{i=1}^4 \frac{\nu_i}{2}\dot{\Psi}_i^2  -\frac{1}{2}E_{J_i}\Psi_i^2 +\frac{E_{J_i}}{24}(B_{ii}\Psi_i +B_{ci}\Psi_c)^4 +  \frac{\nu_c}{2}\dot{\Psi}_c^2  -\frac{1}{2}E_{J_c}\Psi_c^2   + \frac{c_4^{(0)}}{24}\left(\sum_{i=1}^4B_{ic}\Psi_i +B_{cc}\Psi_c\right)^4\\
  &-c_1^{(1)}\cos(\omega_D t)\left(\sum_{i=1}^4B_{ic}\Psi_i +B_{cc}\Psi_c\right)  
  +\frac{c_2^{(2)}}{2}\cos(2\omega_D t)\left(\sum_{i=1}^4B_{ic}\Psi_i +B_{cc}\Psi_c\right)^2 \\
  &+\frac{c_3^{(1)}}{6} \cos(\omega_D t)\left(\sum_{i=1}^4B_{ic}\Psi_i +B_{cc}\Psi_c\right)^3 -\frac{c_4^{(2)}}{24} \cos(2\omega_D t)\left(\sum_{i=1}^4B_{ic}\Psi_i +B_{cc}\Psi_c\right)^4 .
    \end{aligned}
    \label{eq:full_nm_lagrangian}
\end{align}
\end{widetext}
Given that the normal modes of the idle regime with driven amplitude $\delta = 0$ represent the experimentally relevant computational basis, this representation of the Lagrangian allows us to see which interactions between the qubits (represented by the normal modes $\Psi_j$ can be activated. Provided the ramp-up of the drive is slow enough that the idle normal modes at $\delta = 0$ are adiabatically transformed into normal modes at $\delta > 0$, we can estimate the speed of the generated gate operations from the respective pre-factors in Eq. (\ref{eq:full_nm_lagrangian}).

Expanding the normal modes in terms of creation and annihilation operators, one can see that the $c_1^{1}$-terms generate single-photon processes and the $c_2^{2}$-terms generate a parametric two-qubit interaction, when the resonance condition $2\omega_D$ is met. The $c_3^{1}$-terms generate a 3-qubit gate when the resonance condition $\omega_D$ is fulfilled by the applied drive. $c_4^{2}$ generate higher order parametric processes as well as a two-qubit interaction and $c_4^{0}$ renormalizes the anharmonicity.

Using the commutator algebra for bosonic ladder operators, one can find the prefactor for the two-, three- and four-qubit interactions. Where we define the reduced flux in normal modes $\psi = \frac{2\pi}{\phi_0} \Psi$, such that the coefficients \ref{eq:coeff_t-dep} are rewritten as  $\tilde{c}_1^{(1)}  = E_J \left[f_{13} +\alpha f_{1\alpha}\right]$, $\tilde{c}_2^{(2)}= E_J\left[\frac{f_{23}}{3} +\alpha f_{2\alpha}\right]$, $\tilde{c}_3^{(1)} =  E_J \left[\frac{f_{13}}{3^2} +\alpha f_{1\alpha}\right]$, $\tilde{c}_4^{(2)} = E_J \left[\frac{f_{23}}{3^3} +\alpha f_{2\alpha}\right]$ and $\tilde{c}_4^{(0)} = E_J \left[\alpha f_{0\alpha} +\frac{f_{03}}{3^3}\right]$.

For a gate between qubit 1 and qubit 2, as generated by$a_1^\dagger a_2 + \text{h.c.}$, the normal mode calculation yields the prefactor, 
\begin{align*}
    &2B_{1c}B_{2c}\Big(\frac{\tilde{c}_2^{(2)}\cos(2\omega_Dt)}{2}\Big)\psi_1^{\text{zpf}}\psi_2^{\text{zpf}}
    \\
    &+ 12B_{1c}^3B_{2c}\left(\frac{\tilde{c}_4^{(0)}-\tilde{c}_4^{(2)}\cos(\omega_D t)}{24}\right)(\psi_1^{\text{zpf}})^3\psi_2^{\text{zpf}}\\
    &+ 12B_{1c}B_{2c}^3\left(\frac{\tilde{c}_4^{(0)}-\tilde{c}_4^{(2)}\cos(\omega_D t)}{24}\right)\psi_1^{\text{zpf}}(\psi_2^{\text{zpf}})^3 \\
    &+12B_{1c}B_{2c}B_{3c}^2 \left(\frac{\tilde{c}_4^{(0)}-\tilde{c}_4^{(2)}\cos(\omega_D t)}{24}\right)\psi_1^{\text{zpf}}\psi_2^{\text{zpf}}(\psi_3^{\text{zpf}})^2 \\
    &+12B_{1c}B_{2c}B_{4c}^2 \left(\frac{\tilde{c}_4^{(0)}-\tilde{c}_4^{(2)}\cos(\omega_D t)}{24}\right)\psi_1^{\text{zpf}}\psi_2^{\text{zpf}}(\psi_4^{\text{zpf}})^2  \\
    &+12B_{1c}B_{2c}B_{cc}^2 \left(\frac{\tilde{c}_4^{(0)}-\tilde{c}_4^{(2)}\cos(\omega_D t)}{24}\right)\psi_1^{\text{zpf}}\psi_2^{\text{zpf}}(\psi_c^{\text{zpf}})^2.
\end{align*}
which includes contributions from higher order expansion terms that enter due to the normal ordering of the bosonic ladder operators.

For the three-qubit interaction,i $a_1^\dagger a_2a_3 + \text{h.c.}$,in turn  we get the prefactor,
\begin{align*}
    \frac{\tilde{c}_3^{(1)}\cos(\omega_Dt)}{6} \cdot6B_{1c}B_{2c}B_{3c}\psi_1^{\text{zpf}}\psi_2^{\text{zpf}}\psi_3^{\text{zpf}}
\end{align*}

We analyze the values of the above coupling coefficients for the parameters of our system in the next section.

\section{Parameter-analysis for a driven system \label{sec:appendix_driven}}

For choosing the right drive and SNAIL parameters, we do an analysis of the strength of the driven coefficient as a function of the SNAIL parameter $\alpha$ and the drive amplitude $\delta$.
Thus, we show plots for the interactions
\begin{align}
    g_{12} & =\tilde{c}_2^{(2)} B_{1c}B_{2c}\psi_1^{\text{zpf}}\psi_2^{\text{zpf}} \label{eq:g12}\\
    g_{123} & = \tilde{c}_3^{(1)}B_{1c}B_{2c}B_{3c}\psi_1^{\text{zpf}}\psi_2^{\text{zpf}}\psi_3^{\text{zpf}} \label{eq:g123}
\end{align}
as derived in the preceding section, where we only take the leading term into account for the two-qubit interaction $g_{12}$.

\begin{figure}
\centering
        \includegraphics[width = 0.5\textwidth]{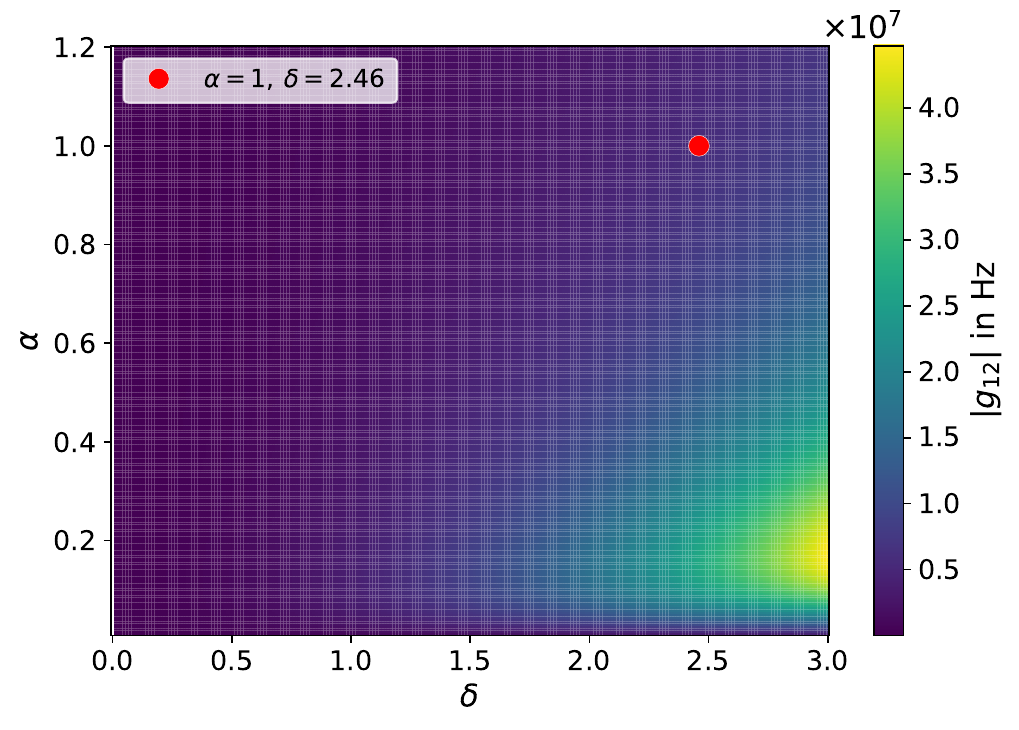}
		    \caption{Absolute value of the coupling coefficient $g_{12}$ as given in Eq. (\ref{eq:g12}). The red dot marks our parameters.}
            \label{fig:g12}
\end{figure}

\begin{figure}
\centering
        \includegraphics[width = 0.5\textwidth]{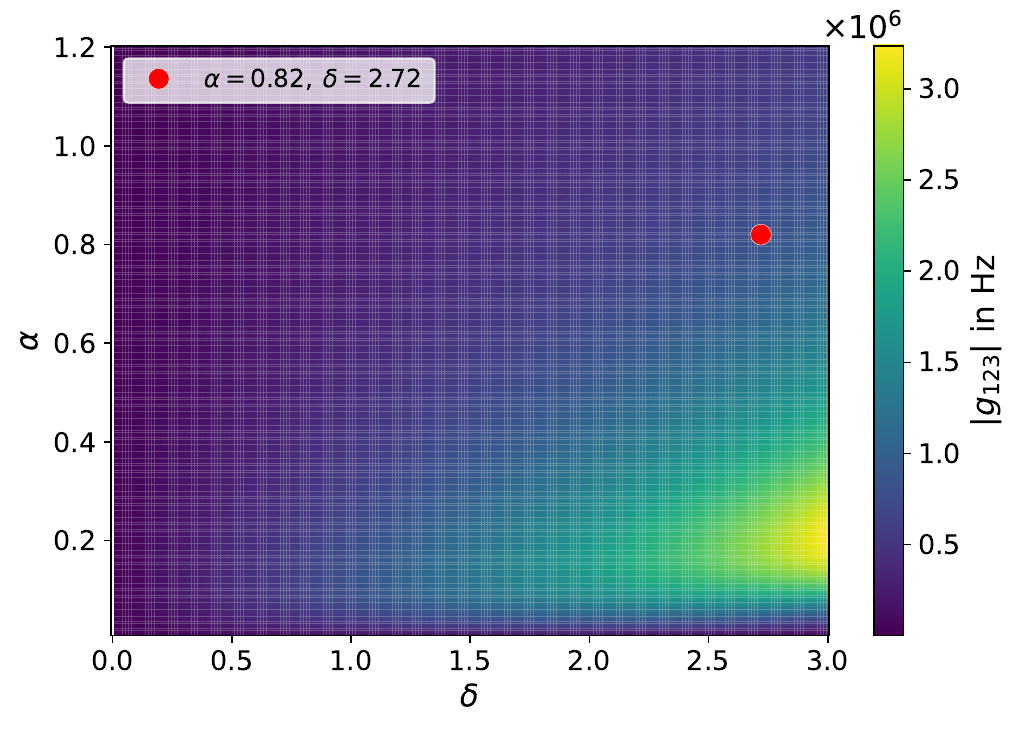}
		    \caption{Absolute value of the coupling coefficient $g_{123}$ as given in Eq. (\ref{eq:g123}). The red dot marks our parameters.}
            \label{fig:g123}
\end{figure}

The resulting plots, shown in Fig. \ref{fig:g12} and Fig. \ref{fig:g123}, allow one to determine which drive amplitude and SNAIL parameters one needs to choose to get a strong coupling strength. We can see for both cases, that the coupling strength is largest when $\alpha \approx 0.3$ and the drive amplitude is large. However, in Appendix \ref{sec:alpha_spread}, we found that $\alpha =1$ is the best choice for hitting the resonance irrespective of the state of the spectator qubits.  

In figure \ref{fig:g12}, we can also see that for the $\alpha$-$\delta$-combination the coupling strength in the normal mode picture would be smaller than for our chosen combination of $\alpha=1$ and a higher drive amplitude, which for the gate times we aim at is ideal.
For the three-qubit gate, we chose a trade-off between a slightly worse spread and stronger coupling in order to achieve a faster gate. Thus, we selected $\alpha =0.82$ and $\delta=2.72$

From these values one can estimate the gate times as $t_{\text{gate}}= \frac{\pi}{4 g}$, where $g = g_{12}$ for the $\sqrt{\text{iSWAP}}$-gate and $g = g_{123}$ for the three-qubit gate.
From the  values in the plots, we expect gate times of 46.6 ns for the two-qubit gate and 308.4 ns for the three-qubit gate. In the simulations of the gates we found gate times of 48 ns for the two-qubit gate in figure \ref{fig:snail12} and 200 ns for the three-qubit gate in figure \ref{fig:3q}. Considering the rise and fall time of the super-Gaussian pulse, the analysis fits the two-qubit gate well, for the three-qubit gate however, other processes like higher orders are most likely contributing to the faster gate time.

\section{System parameters}
The parameters used in the simulations for the SQUID and SNAIL system can be found in the following tables \ref{table:table_squid_capE}, \ref{table:table_snail_capE}.

\begin{table}[h]
\centering
\begin{tabular}{ |c|c|c| } 
 \hline
  & Capacity [fF] & Josephson energy [GHz]\\ 
  \hline
Qubit 1 & 94.0 & 35.0 \\ 
Qubit 2 & 94.0 & 30.0 \\ 
Qubit 3 & 94.0 & 26.5 \\ 
Qubit 4 & 94.0 & 21.5 \\ 
SQUID   & 86.0 & 65.0 \\ 
 \hline
\end{tabular}
\caption{SQUID-coupler parameters, the coupling capacitance is  8.0 fF.}
\label{table:table_squid_capE}
\end{table}

\begin{table}[h]
\centering
\begin{tabular}{ |c|c|c| } 
 \hline
  & Capacity [fF] & Josephson energy [GHz]\\ 
  \hline
Qubit 1 & 85.0 & 29.0 \\ 
Qubit 2 & 85.0 & 27.0 \\ 
Qubit 3 & 85.0 & 24.0 \\ 
Qubit 4 & 85.0 & 20.0 \\ 
SNAIL   & 40.0 & 80.0 \\ 
 \hline
\end{tabular}
\caption{SNAIL-coupler parameters, the coupling capacitance is 9.5 fF.}
\label{table:table_snail_capE}
\end{table}

\section{Fidelities for different gate times}
\begin{figure}[h]
\centering
        \includegraphics[width = 0.5\textwidth]{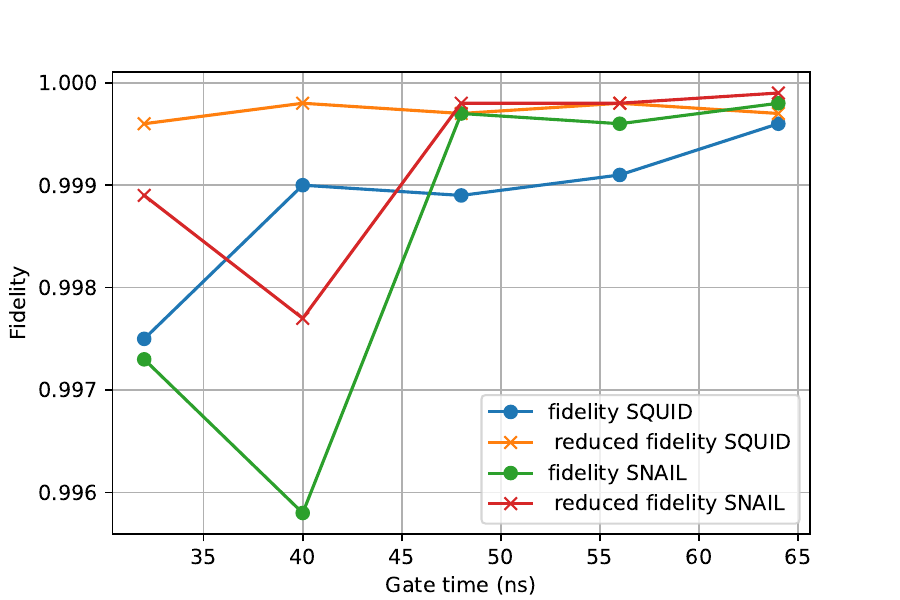}
		    \caption{Different fidelities and reduced fidelities for an $\sqrt{\text{iSWAP}}$-gate between qubits 1 and 2 for different gate times.}
            \label{fig:fidelities}
\end{figure}

We plot the fidelities for different gate times in figure \ref{fig:fidelities}, in some cases there could be some improvement in the optimization, still the trend shows that the fidelity and reduced fidelity improve with higher gate times (and smaller amplitudes).

\nocite{*}

\bibliography{mybib}

\end{document}